%% file: template.tex
\documentclass{article}

\PassOptionsToPackage{numbers, compress}{natbib}
\usepackage[preprint]{neurips_2026}

\usepackage{upquote}
\usepackage[utf8]{inputenc}
\usepackage[T1]{fontenc}
\usepackage{hyperref}
\usepackage{url}
\usepackage{booktabs}
\usepackage{amsfonts}
\usepackage{nicefrac}
\usepackage{microtype}
\usepackage{xcolor}
\usepackage{wrapfig}
\usepackage{cutwin}
\usepackage{lipsum}
\usepackage{caption}
\usepackage{subcaption}
\usepackage{CJK}
\usepackage{multirow}
\usepackage[absolute,overlay]{textpos}
\usepackage{graphicx}
\usepackage{tabularx}
\usepackage{makecell}
\usepackage{pifont}
\usepackage{caption}
\usepackage[table]{xcolor}
\newcommand{\cmark}{\textcolor{green!60!black}{\ding{51}}}
\newcommand{\xmark}{\textcolor{red}{\ding{55}}}
\newcommand{\pmark}{\textcolor{orange!80!black}{\scalebox{1.5}{$\circ$}}}
\newcolumntype{Y}{>{\centering\arraybackslash}X}

\renewcommand{\arraystretch}{1.15}
\usepackage{arydshln}
\input{tool}

\usepackage{tcolorbox}
\tcbuselibrary{breakable,skins}
\usepackage{enumitem}

\usepackage{tikz}
\usetikzlibrary{shapes.geometric, arrows.meta, positioning, calc, backgrounds, shadows}
\usepackage{xcolor}

\definecolor{boxblue}{RGB}{230, 242, 250}
\definecolor{boxorange}{RGB}{255, 245, 230}
\definecolor{boxgreen}{RGB}{235, 247, 240}
\definecolor{titleblue}{RGB}{70, 130, 180}
\definecolor{titleorange}{RGB}{210, 140, 50}
\definecolor{icbcbg}{RGB}{245, 245, 245}

\title{%
ICBCBench: An Industry Consortium Benchmark for Financial Deep Research
}

\author{
\parbox{\textwidth}{
\normalfont
\raggedright
\textbf{Authors}\\
Weiya Li$^{1,\ddagger}$\thanks{Core authors. $^{\ddagger}$ Project leader. Email: weiyali126@outlook.com, zwtang1220@gmail.com, jonah\_he@163.com} \quad
Zhiwei Tang$^{4,*}$ \quad
Yizhou He$^{3,*}$ \quad
Chenghao Wang$^{1}$ \quad 
Liang Feng$^{3}$ \quad
Xiao Sun$^{2}$ \vspace{0.5mm}
Dongrui Liu$^{5}$ \quad
Zichen Wen$^{2}$ \quad 
Hu Wei$^{6}$ \quad 
Jinghang Wang$^{6}$ \quad 
Yi Luo$^{1}$\thanks{Corresponding authors. Email: zhanglinfeng@sjtu.edu.cn, guo\_li@fudan.edu.cn} \quad
Li Guo$^{3,36\dagger}$ \quad\\
Linfeng Zhang$^{2,\dagger}$ 
\vspace{2.5mm}\\
\textbf{Contributors (Alphabetical Order)}\\
Anping Liu$^{1}$,
Armstrong Sheng Chen$^{8}$,
Baolong Liu$^{9}$,
Bing Huang$^{1}$,
Bingxun Li$^{3}$,
Bingyan Yang$^{1}$,
Boyan Chen$^{1}$,
Chen Xia$^{24}$,
Chengyan Liu$^{1}$,
Chenyang Zhang$^{27}$,
Chunhui Zhang$^{11}$,
Dalong Kuang$^{1}$,\\
David Lee Kuo Chuen$^{7,22,33,38}$,
Dong Shen$^{1}$, 
Dongqing Cao$^{42}$,
Fengting Li$^{23}$,
Haiming Zhao$^{41}$, \\
Hanchen Wang$^{14}$,
Hongjun Huang$^{16}$,
Hongsheng Gao$^{1}$,
Hongyu Yao$^{1}$, 
Huajie Liu$^{1}$,
Huajing Si$^{1}$,\\
Huiping Ma$^{1}$,
Hui Li$^{37}$,
Huanxiang Song$^{10}$,
Jialiang Chen$^{1}$,
Jiankang Wang$^{14}$, 
Jian Lu$^{1}$,
Jiaqi Li$^{30}$,\\
Jiawei Li$^{39}$,
Jiaxi Wang$^{1}$,
Jiliu Xia$^{25}$,
Jingnan Cao$^{34}$,
Jingyi Shen$^{3}$, 
Jinzhi Xu$^{12}$,  
Junshuo Jia$^{3}$,\\
Le Chang$^{1}$,
Li Kang$^{1}$, 
Lily Li$^{31}$, 
Ling Ni$^{1}$,  
Liuyin Chen$^{43}$,
Lunsheng Song$^{1}$, 
Luyuan Zhao$^{19}$,\\
Mengdi Zhang$^{1}$,
Minghui Su$^{28}$, 
Naping Li$^{17}$,
Peng Yuan$^{26}$,
Rui Zhu$^{15}$,
Rurui Yang$^{29}$, 
Shan Zhong$^{13}$, \\
Shaohua Han$^{1}$,
Shihao Zhou$^{3}$, 
Shizhen Kang$^{18}$,
Weijie Chen$^{3}$,
Weiliang Yao$^{40}$,
Wencheng Xu$^{19}$,\\
Xiangchen Kong$^{3}$, 
Xiang Li$^{1}$,
Xiangyang Qin$^{3}$, 
Xiaobin Zhang$^{3}$, 
Xin Jin$^{1}$, 
Xinyue Shao$^{3}$,
Yang Li$^{1}$,\\
Yameng Zhou$^{1}$,
Yang Shen$^{28}$,
Yayao Jiang$^{20}$,
Yichao Huang$^{3}$,
Yikun Yin$^{3}$, 
Yu Wang$^{22}$,
Yuanli Wen$^{2}$,\\
Yue Qin$^{32}$, 
Yufei Shi$^{1}$,
Yujie Zha$^{17}$,
Yuting Guang$^{3}$,  
Yuzhou Hu$^{35}$,
Zhe Wang$^{1}$,
Zhiyuan Xia$^{1}$,\\
Zhiyuan Xu$^{14}$, 
Zhonghuan Wang$^{1}$,
Zixin Wei$^{3}$, 
Zixun Zheng$^{21}$
\vspace{2.5mm}\\
\textbf{Affiliations}\\
$^{1}$ Industrial and Commercial Bank of China, 
$^{2}$ Shanghai Jiao Tong University, \\
$^{3}$ School of Economics, Fudan University, 
$^{4}$ East China Normal University, \\
$^{5}$ Shanghai AI Laboratory, 
$^{6}$ Alibaba Group,
$^{7}$ Asia Pacific Exchange, \\
$^{8}$ Beijing Dacheng Law Offices, LLP (Shanghai)
$^{9}$ Beijing FinTech Industry Alliance,\\
$^{10}$ BOC International (China), 
$^{11}$ CCX Heyi Information Technology (Shanghai),\\
$^{12}$ Changjiang Securities, 
$^{13}$ Chengtong Securities, 
$^{14}$ China Development Bank,\\
$^{15}$ Chuanshan Fund, 
$^{16}$ CIB Wealth Management,
$^{17}$ CICC,
$^{18}$ CITIC Futures, 
$^{19}$ CITIC Securities,\\
$^{20}$ E Fund Management,
$^{21}$ Fofund Fund Distribution,
$^{22}$ Global FinTech Institute, \\
$^{23}$ Guosheng Securities,
$^{24}$ Guolian Minsheng Securities Underwriting and Sponsorship,\\
$^{25}$ Guotai Haitong Securities,
$^{26}$ Hongnuo Venture Capital (Shenzhen),
$^{27}$ Huachuang Securities,\\
$^{28}$ Huatai Securities, 
$^{29}$ J Trust Global Securities, 
$^{30}$ JF SmartInvest,
$^{31}$ Man Group,\\
$^{32}$ Nanyang Commercial Bank,
$^{33}$ National University of Singapore,
$^{34}$ Orient Securities,\\
$^{35}$ Ping An Asset Management,
$^{36}$ Shanghai Institute of International Finance and Economics,\\
$^{37}$ Sealand Securities,
$^{38}$ Singapore University of Social Sciences, 
$^{39}$ Southwest Securities,\\
$^{40}$ Ubiquant Investment,
$^{41}$ University of Birmingham,
$^{42}$ Value Partners Group Limited,\\
$^{43}$ Zhejiang Zheyin Financial Leasing
}
}

\begin{document}

\maketitle

\footnotetext[1]{Including the Big Data \& AI Lab, the Modern Finance Research Institute of ICBC, the Global Markets Department, the Private Banking Department, and ICBC Wealth Management Co., Ltd.}

\thispagestyle{plain}

\newpage
\begin{abstract}
\input{abstract}
\end{abstract}

\input{main}



\end{document}

%% file: tool.tex
\usepackage{enumitem}
\usepackage{verbatim}
\usepackage{xspace}
\usepackage{diagbox}
\usepackage{array}

\usepackage{multirow}
\usepackage{makecell}
\usepackage{diagbox}
\usepackage{url}
\usepackage{comment}
\usepackage{booktabs}
\usepackage{amsmath}
\usepackage{scalerel}
\usepackage[figuresright]{rotating}
\usepackage{threeparttable}
\usepackage{algorithm}
\usepackage{algorithmic}
\usepackage{siunitx}
\usepackage{stackengine}
\usepackage{pifont}
\usepackage{tabularx}
\usepackage{listings}

\usepackage{tcolorbox}
\definecolor{grey}{RGB}{128,128,128}

\usepackage{mathtools}
\usepackage{multirow}
\usepackage{multicol}
\usepackage{booktabs}

\usepackage{amsmath,amsfonts,bm}

\def\eqref#1{equation~\ref{#1}}

\def\1{\bm{1}}

\DeclareMathAlphabet{\mathsfit}{\encodingdefault}{\sfdefault}{m}{sl}
\SetMathAlphabet{\mathsfit}{bold}{\encodingdefault}{\sfdefault}{bx}{n}

\definecolor{mygreen}{rgb}{0.7, 1.0, 0.7}
\definecolor{myblue}{rgb}{0.7, 0.8, 1.0}

\definecolor{promptcolor}{RGB}{90,90,160} 

%% file: abstract.tex
With the rapid advancement of Deep Research Agents in knowledge-intensive domains such as finance, establishing reliable and domain-aligned evaluation standards remains a critical challenge. Existing benchmarks focus on either closed-ended question answering or open-ended report evaluation, failing to jointly capture retrieval–reasoning accuracy and end-to-end research quality required in real-world workflows. We introduce ICBCBench, a consortium-driven benchmark for financial deep research, developed in collaboration with domain experts from a broad range of financial institutions and academia, involving over 50 experts across more than 40 organizations. It adopts a dual-track paradigm integrating objective tasks with verifiable answers and subjective long-form report evaluation, enabling complementary assessment of retrieval–reasoning accuracy and end-to-end report quality in terms of expert alignment, citation consistency, and source quality. Experiments on state-of-the-art DRAs and large language models reveal substantial gaps in complex reasoning, factual grounding, and report quality, highlighting the challenges of achieving industry-level performance. Our dataset and evaluation framework are available at \url{https://github.com/DeepFin-Intelligence/ICBCBench}.


%% file: main.tex
\section{Introduction}
\vspace{-1pt}
With the rapid advancement of large language models (LLMs), agent-based paradigms for automating complex tasks are emerging as a key application trend in industry. Among these, Deep Research Agents (DRAs) have become a representative direction~\cite{openai2024deepresearch,google2024geminideepresearch,perplexity2025deepresearch}. Given a user query, a DRA can execute a full research pipeline, including problem analysis, research planning, iterative retrieval and reasoning, and the integration of information from diverse sources, ultimately producing a structured report comparable to that of professional research analysts. This paradigm is particularly well-suited to domains such as finance, with strong demand for large-scale research report generation, offering significant practical value and broad application potential.

Despite rapid progress, evaluating financial research reports remains highly challenging. 
Financial research is time-sensitive, knowledge-intensive, and risk-sensitive, requiring reliable sources, rigorous analysis, factual accuracy, and deep reasoning to support high-stakes decision-making.
Prior work has explored several evaluation paradigms for deep research. 
Benchmarks such as HLE, GAIA, BrowseComp, DeepSearchQA, and FinSearchComp~\cite{phan2025humanity,mialon2023gaia,Wei2025BrowseCompAS,Gupta2026DeepSearchQABT,Hu2025FinSearchCompTA} use short, verifiable closed-ended questions to indirectly assess retrieval and reasoning. 
Another line of work, represented by DeepResearch Bench, ResearchRubrics, LiveResearchBench, DRACO, and related benchmarks~\cite{Du2025DeepResearchBA,Yao2025DrBA,wang2025liveresearchbenchlivebenchmarkusercentric,Sharma2025ResearchRubricsAB,Zhong2026DRACOAC,wan2025deepresearcharenaexamllms,abaskohi2026drbenchrealisticbenchmarkenterprise}, evaluates long-form reports through citation accuracy and checklist-based expert assessment. 
A comparison of representative benchmarks is provided in Appendix Table~\ref{tab:benchmark_comparison}.

Nevertheless, existing approaches remain insufficient for the financial domain. Most benchmarks rely on a \textbf{single evaluation paradigm}, focusing either on objective question answering or subjective report assessment, which captures only a partial view of financial research capability. Objective evaluation offers scalable and reproducible measurement of factual correctness and reasoning accuracy, but is limited in assessing open-ended analytical quality. Subjective evaluation better reflects real-world analytical tasks, yet it is harder to standardize and reproduce across evaluators. These limitations motivate a dual-track evaluation design, analogous to financial qualification exams such as CPA~\cite{cicpa_official}, ACCA~\cite{acca_official}, and CFA~\cite{cfa_institute}, where objective and subjective components are jointly used to assess both correctness and higher-order reasoning.

Beyond evaluation limitations, existing financial report benchmarks also suffer from \textbf{impractical  task design}, making them poorly aligned with real-world research workflows. Many prompts are overly broad, lack comparability across samples, and weakly reflect practical financial research needs~\cite{Du2025DeepResearchBA,Yao2025DrBA,wang2025liveresearchbenchlivebenchmarkusercentric,wan2025deepresearcharenaexamllms,Sharma2025ResearchRubricsAB} (e.g., \textit{"Researching how the world's wealthiest governments invest"} or \textit{"What are the investment philosophies of Duan Yongping, Warren Buffett, and Charlie Munger?"}). Such open-ended queries often require iterative refinement with domain experts before yielding usable outputs, reducing efficiency and hindering reliable, domain-informed evaluation.

To address these challenges, we propose ICBCBench, an industry consortium benchmark for financial deep research. ICBCBench is developed through large-scale collaboration with domain experts from financial institutions and academia, involving over \textbf{50} analysts and researchers across more than \textbf{40} organizations, spanning banks, securities firms, and asset managers. Most participants have at least \textbf{10 years} of combined academic and professional experience in finance-related fields.
Subjective tasks are designed and validated by experienced industry professionals to reflect real-world research practice, while objective tasks are constructed with support from well-trained students in economics and finance. A key design of ICBCBench is a structured task formulation for subjective report evaluation. Each task is defined as a concrete research assignment with \textbf{2–5} explicit analytical objectives and validated by at least \textbf{3} domain experts, significantly improving clarity, comparability, and alignment with real-world financial analysis.

In addition, ICBCBench adopts a unified dual-track evaluation framework that integrates objective questions with verifiable ground-truth answers and subjective report-generation tasks, enabling joint assessment of retrieval–reasoning accuracy and long-form analytical quality, as illustrated in Figure~\ref{fig:icbcbench_pipeline}. The benchmark comprises \textbf{120 tasks}, including 60 in English targeting global markets and 60 in Chinese targeting domestic scenarios. This market-aware design reflects real-world financial applications~\cite{Hu2025FinSearchCompTA}. For each language, ICBCBench includes \textbf{40} objective questions and \textbf{20} subjective tasks, enabling comprehensive evaluation across factual accuracy, citation grounding, and high-level reasoning.
Together, these designs enable a comprehensive evaluation framework that jointly captures retrieval–reasoning accuracy and end-to-end research quality, aligning closely with real-world financial practice. Table~\ref{tab:affiliation_distribution} summarizes representative participating institutions, reflecting the strong industry and academic collaboration behind ICBCBench.

\input{figures/main_figure}

In summary, our contributions are as follows:

\noindent\begin{enumerate}[leftmargin=*, itemsep=4pt, topsep=0pt, parsep=0pt]

\item \textbf{A consortium-driven, industry-aligned benchmark for financial deep research.} 
We introduce ICBCBench, a consortium-driven benchmark covering diverse real-world scenarios across capital markets, banking, and insurance, ensuring high task quality, comparability, and strong alignment with professional financial workflows.

\item \textbf{A unified dual-track evaluation paradigm.} 
We propose a framework that integrates objective questions with verifiable ground-truth answers and subjective report-generation tasks, enabling joint assessment of retrieval–reasoning accuracy and long-form analytical quality.

\item \textbf{An expert-aligned hybrid evaluation framework.} 
We develop a hybrid methodology combining prompt-based expert rubrics, citation consistency checking, and source quality assessment, providing a multi-dimensional evaluation protocol grounded in real-world financial practice.

\end{enumerate}

\section{Dataset}
This section presents the construction of ICBCBench, including its industry-grounded data sources, quality control process, task statistics and taxonomy, public/private splits, and the design of objective and subjective tasks.

\textbf{Data Sources and Quality Control.} All tasks in ICBCBench are grounded in real-world financial research needs from industry. To ensure data quality and realism, we adopt a rigorous multi-stage construction pipeline consisting of source curation, task design, LLM-based validation, expert cross-review, and finalization. During task construction, subjective report-based tasks were developed by over \textit{50} researchers and analysts from financial institutions worldwide, most with over three years of domain experience, while objective questions were designed with support from \textit{15} well-trained students in economics and finance. Appendix Table~\ref{tab:affiliation_distribution} presents the institutional affiliations of selected contributors. To further standardize task design, we establish finance-specific authoring principles covering accuracy and compliance, domain relevance, task depth and complexity, and scope diversity, as detailed in Appendix Table~\ref{tab:authoring_principles}.

\textbf{Task Statistics and Taxonomy.}
ICBCBench contains 180 tasks in total, including 120 publicly released tasks and a private hold-out set of 60 tasks for hidden evaluation. The complete benchmark spans 4 primary financial domains and 34 secondary sub-domains, while the public subset covers 27 sub-domains. The public set includes 60 English tasks targeting global markets and 60 Chinese tasks targeting domestic scenarios, supporting reproducible evaluation and leaderboard construction. Its distribution across task types and domains is shown in Figure~\ref{fig:task_taxonomy}, and the complete taxonomy is provided in Appendix Table~\ref{tab:financial_taxonomy}.

\input{figures/task_taxonomy}

For each language in the public subset, ICBCBench includes 40 objective tasks and 20 subjective tasks. Objective tasks assess retrieval and reasoning accuracy through verifiable answers, while subjective tasks evaluate analytical depth, logical coherence, citation grounding, and presentation quality. Representative examples are shown in Appendix Figures~\ref{fig:appendix_case_objective} and~\ref{fig:appendix_case_subjective}.

\subsection{Objective Tasks}
Inspired by HLE~\cite{phan2025humanity} and BrowseComp~\cite{Wei2025BrowseCompAS}, our objective tasks are designed to assess verifiable financial retrieval and reasoning. 
They use three answer formats, including multiple-choice questions, numerical computation, and short phrase matching. Each question is annotated with required tools such as Search, Visit, Multi-modality, and Coding, as well as a difficulty level from Level 1 to Level 3 based on source count and reasoning complexity, with detailed criteria provided in Appendix~\ref{subsec:appendix_difficulty}. To ensure real-world grounding and verifiability, we construct objective questions from over 20,000 financial research reports through a five-stage pipeline, including initial task authoring, LLM-based screening, human solving and cross-review, task refinement and candidate selection, and final acceptance review. 
Details of the construction pipeline are provided in Appendix~\ref{subsec:objective_task_details}.

\subsection{Subjective Tasks}
Subjective tasks are derived from three primary sources: enterprise DeepResearch dialogues, authentic financial reports, and expert-designed tasks. These sources jointly capture real-world user needs, representative report structures, and emerging topics identified by domain experts.

\textbf{Enterprise dialogue tasks.}
We collect raw user queries $\mathcal{Q} = \{q_i\}_{i=1}^{N}$ from internal DeepResearch applications within financial institutions, where $N = 6{,}376$, spanning domains such as capital markets, banking, and insurance. Although these queries reflect real-world user needs, their quality is highly uneven and many are unsuitable for direct use as benchmark tasks. We therefore manually curate approximately $100$ representative seed queries and refine them into structured research prompts using GPT-5.4~\cite{openai2025gpt54} with curated exemplars. The refinement process converts colloquial, overly broad, or underspecified queries into concrete research assignments with explicit background, constraints, and output structure. Details are provided in Appendix~\ref{subsec:subjective_task_details}.

\textbf{Report-derived and expert-designed tasks.}
Beyond enterprise dialogue data, we invited industry financial experts to recommend representative research reports and supplemented underrepresented topics using our collected report corpus. This process yielded 100 representative reports, from which GPT-5.4 extracted core research themes and reformulated them into structured report analysis tasks. Because financial research is highly time-sensitive, we further asked domain experts to refine existing tasks and contribute additional real-world research tasks, yielding 18 high-quality expert-designed tasks.

\noindent\textbf{Final Subjective Task Set.}
From the three sources above, we construct 218 candidate subjective tasks and assign each task to at least three domain-specific financial experts for review, scoring, and feedback. Based on expert feedback and recommendation scores, we select 40 tasks as the final subjective task set. These tasks cover three primary domains, including capital markets, banking, and insurance. Appendix Figure~\ref{fig:appendix_case_subjective} presents representative subjective tasks.

\section{Evaluation Methodology}
ICBCBench adopts a dual-track evaluation methodology tailored to financial deep research. Objective tasks are scored by answer correctness against verifiable ground-truth answers, with confidence calibration reported as an auxiliary metric. 
Subjective reports are assessed through an expert-aligned framework combining rubric-based evaluation, citation consistency, and source quality verification.

\subsection{Objective Task Evaluation}
Following HLE~\cite{phan2025humanity}, we adopt an LLM-as-a-judge pipeline for objective evaluation. Models are prompted to output a reasoning process, a final answer, and a confidence score ranging from 0\% to 100\%. We use GPT-5.4~\cite{openai2025gpt54} as the judge model with structured decoding to parse each response into four fields: \textit{extracted\_final\_answer}, \textit{reasoning}, \textit{correct}, and \textit{confidence}. For numerical computation tasks, equivalent representations and small tolerances are allowed, while short phrase matching and multiple-choice questions require exact equivalence. The primary objective score is then computed as the overall proportion of correctly answered questions, scaled to a standard 0--100 point range.

In financial research, reliable uncertainty estimation is critical because overconfident incorrect predictions may lead to high-risk decisions. 
We evaluate calibration using Root Mean Square Calibration Error (RMSCE), which measures the discrepancy between predicted confidence and empirical accuracy:
\begin{equation}
    \text{RMSCE} = \sqrt{ \sum_{k=1}^{K} \frac{|S_k|}{N} \left( \text{acc}(S_k) - \text{conf}(S_k) \right)^2 } .
\end{equation}
The $N$ samples are sorted by confidence and partitioned into $K$ equal-size bins, with 10 samples per bin. Here, $S_k$ denotes the samples in bin $k$, while $\text{acc}(S_k)$ and $\text{conf}(S_k)$ denote empirical accuracy and average confidence, respectively. 
The solver and judge prompts are provided in Appendix Figures~\ref{fig:solver_prompt} and~\ref{fig:judge_prompt}, and model versions are listed in Appendix Table~\ref{tab:model_versions}.

\subsection{Subjective Task Evaluation}
Subjective reports are evaluated through three components: task-specific expert rubrics as the primary measure of analytical quality, and citation consistency and source quality as auxiliary signals for factual grounding, traceability, and reliability.

\textbf{Expert Rubric Evaluation.}
For each subjective task, we design a task-specific rubric tailored to the corresponding financial report type. Each rubric follows a 100-point scale and typically consists of 4--6 high-level dimensions and 12--16 fine-grained sub-dimensions, yielding an expert score $S_{\text{expert}} \in [0,100]$ by aggregating scores across all dimensions. To ensure expert alignment, rubrics are drafted by financial experts, refined with GPT-5.4~\cite{openai2025gpt54}, reviewed by at least three additional domain experts, and finalized by the organizers. During evaluation, Gemini-3.1-Pro-Preview~\cite{googledeepmind2025gemini31pro} is used as the LLM judge and instructed to score each report according to the task-specific rubric, with justifications grounded in the corresponding sub-dimensions. Additional details are provided in Appendix~\ref{subsec:rubric_construction_details}, with an example rubric and judge prompt shown in Appendix Figures~\ref{fig:appendix_case_expert_rubric} and~\ref{fig:subjective_judge_prompt}.

\textbf{Citation Consistency Checking.}
Following the FACT-style evaluation in DeepResearch Bench~\cite{Du2025DeepResearchBA}, we extract and deduplicate statement--URL pairs from each report and judge whether each cited source supports the corresponding statement. For task $t$, let $U_t$ denote the set of deduplicated statement--URL pairs and $N_{s,t}$ the number of supported pairs. 
The citation score is defined as:
\begin{equation}
S_{\text{citation}}^{(t)} =
\begin{cases}
\frac{N_{s,t}}{|U_t|}, & |U_t| > 0 \\
0, & |U_t| = 0
\end{cases}
\end{equation}
Failed or inaccessible citations are treated as unsupported. The overall citation score is averaged across all tasks and scaled to $[0,100]$:
$
S_{\text{citation}} = \frac{100}{T} \sum_{t=1}^{T} S_{\text{citation}}^{(t)},
$
where $T$ is the number of subjective tasks.

\textbf{Source Quality Verification.}
We assess source quality along two dimensions: authority and timeliness. Authority reflects the credibility of cited sources, such as official institutions, financial institutions, and major financial media outlets, while timeliness measures information recency. The resulting source score $S_{\text{source}} \in [0,100]$ is computed as detailed in Appendix~\ref{subsec:source_scoring}.

\textbf{Final Scoring Function.}
The final subjective score assigns 80\% weight to expert-aligned rubric evaluation and 10\% each to citation consistency and source quality:
\begin{equation}
S = 0.8 \cdot S_{\text{expert}} + 0.1 \cdot S_{\text{citation}} + 0.1 \cdot S_{\text{source}}
\end{equation}
This weighting prioritizes expert-aligned analytical quality while incorporating citation and source-based checks for factual grounding and reliability.

\section{Experiments and Analysis}
We evaluate a diverse set of state-of-the-art Deep Research Agents (DRAs) and general-purpose large language models across both closed-source systems and open-source frameworks, as detailed in Appendix~\ref{subsec:appendix_model_list}. Detailed configurations of all evaluated models and frameworks, including backbone models and version information, are provided in Table~\ref{tab:model_versions}.

\input{tables/main_results}

\subsection{Main Results}
\noindent\textbf{Overall Performance.} Table~\ref{tab:main_results} shows that open agentic frameworks are highly competitive with, and in several cases outperform, closed-source Deep Research systems. In the Global (EN) scenario, DeerFlow(+GPT-5.5) achieves the highest overall score (58.67), followed by Gemini-deep-research (57.38). In the Chinese (ZH) scenario, OpenClaw(+GPT-5.5) ranks first with an overall score of 63.38. Among closed-source systems, Gemini-deep-research is the most robust, achieving the second-highest overall score in both EN and ZH tracks.

\noindent\textbf{Objective vs.\ Subjective Performance Gap.}
A consistent pattern in Table~\ref{tab:main_results} is that most systems perform substantially better on Subjective tasks than on Objective tasks. While many models achieve Subjective scores above 50.00, only a few systems, mostly open agentic frameworks, exceed 50.00 on Objective tasks. This gap suggests that precise, verifiable financial reasoning remains more challenging than long-form report generation, even for strong Deep Research systems.

\input{figures/score_comparison_butterfly_chart}

\noindent\textbf{Cross-Lingual Discrepancies.}
Figure~\ref{fig:score_comparison_butterfly_chart} reveals substantial cross-lingual variation across systems. Many models exhibit positive localization gaps ($\Delta$), indicating stronger performance on the Global (EN) track, whereas several open agentic frameworks show negative gaps and stronger adaptation to Chinese financial scenarios. 
OpenClaw(+GPT-5.5) presents the clearest Chinese-oriented pattern, improving from 54.80 in EN to 63.38 in ZH, while Gemini-deep-research and DeerFlow(+GPT-5.5) show the most balanced overall performance across languages.

Table~\ref{tab:main_results} further indicates that these gaps are often driven by dimension-level imbalances rather than uniform performance shifts. 
For example, OpenAI-o3-deep-research achieves the highest EN Subjective score (71.84), but its ZH performance is limited by a much lower Objective score (32.50). 
These results suggest that robust financial Deep Research systems must maintain balanced capabilities across languages, objective reasoning, and subjective report generation.

\subsection{Human Consistency}
Given the highly subjective nature of financial report evaluation, we designed a series of consistency experiments comparing human experts and LLM judges to validate the effectiveness of our proposed Expert Rubrics.

\textbf{Human Expert Data Collection.} Reading and evaluating long-form financial reports presents significant professional barriers and demands substantial time commitments. To address this, we sampled reports generated by five representative DeepResearch Agents across our 60 subjective questions. These were distributed to more than 30 financial experts from various institutions, the majority of whom were analysts or researchers with over three years of industry experience. Each expert was asked to select and score up to 5 questions strictly within their domain of expertise.

\textbf{Quality Control.} Human inconsistencies can unfairly penalize LLM evaluation. Following DeepResearch Bench~\cite{Du2025DeepResearchBA}, we measure inter-rater reliability using the \textbf{Intraclass Correlation Coefficient (ICC)}. Samples indicating poor human consensus ($\text{ICC} < 0$) were rigorously excluded. This yielded a high-quality dataset of 36 evaluation samples from 25 experts across 15 questions (5 English, 10 Chinese), each validated by at least two experts with strong consensus.

\textbf{Evaluation Metrics.} To comprehensively evaluate the alignment between LLM judges and human experts on this filtered subset, we establish a robust evaluation framework utilizing the following complementary metrics, including Spearman's Rank Correlation Coefficient ($\rho$), Mean Absolute Error (MAE), and Pairwise Agreement Rate (PAR), whose details are introduced in Appendix~\ref{subsec:append_evaluation_detail}.

\input{human_consistency/human_consistency}
\input{human_consistency/overall_consistency}

\noindent\textbf{Relative Ranking and Pairwise Preferences.}
We first evaluate the comparative judgment capabilities using Spearman's $\rho$ and the Pairwise Agreement Rate (PAR). As detailed in Table~\ref{tab:overall_consistency}, human experts establish an empirical consensus ceiling with an inter-expert $\rho$ of 0.638 and a PAR of 0.711. Remarkably, the \textbf{Expert-LLM alignment matches and even slightly exceeds this human baseline}. Figure~\ref{fig:human_consistency_rank} illustrates this robust alignment across diverse tasks, yielding an overarching Expert-LLM $\rho$ of 0.643. Furthermore, in binary decision-making, the LLM judges agree with human experts in 72.9\% of pairwise comparisons (PAR=0.729), outperforming the natural agreement rate among humans themselves, suggesting that our Expert Rubrics effectively distill complex financial reasoning into reproducible machine directives, enabling LLMs to serve as highly reliable comparative judges.

\noindent\textbf{Absolute Scoring Deviation and Stability.}
While ranking reflects relative preferences, we employ Mean Absolute Error (MAE) to evaluate the systemic deviation in absolute scoring. Figure~\ref{fig:human_consistency_score} establishes the inherent variance among human evaluators, with an inter-expert MAE of 15.36 points. Strikingly, Table~\ref{tab:overall_consistency} reveals that the absolute score deviation between LLM judges and human experts is lower at 12.20 points. More importantly, the internal deviation among different advanced LLMs (Inter-LLM) is exceptionally minimal (5.83 pts). This significant contrast (15.36 vs. 5.83 pts) convincingly demonstrates that LLMs, guided by our structured rubrics, effectively transcend individual human subjectivity and fatigue. They provide a \textit{super-human scoring stability} for open-ended financial tasks, free from the scale-drifting often observed in human evaluations.

\subsection{Traditional Deep Research vs.\ Open-Agentic Paradigms}
\noindent\textbf{Framework Gains and Backbone Bottlenecks.}
Table~\ref{tab:main_results} shows that open-agentic frameworks (e.g., OpenClaw~\cite{openclaw2026openclaw}, DeerFlow~\cite{bytedance2026deerflow}) can bring substantial performance gains and, in several cases, outperform monolithic closed-source systems. For instance, deploying GPT-5.5 within DeerFlow improves the Overall EN score from 45.09 to 58.67, while OpenClaw raises the Overall ZH score from 42.41 to 63.38. Notably, Figure~\ref{fig:score_comparison_butterfly_chart} suggests that frameworks may alter the cross-lingual behavior of their backbone models. We hypothesize that the pronounced Chinese-oriented pattern ($\Delta$) observed in certain configurations may be related to the adaptation of localized toolchains, where external retrieval APIs or data parsing skills are better suited to Chinese financial corpora, thereby amplifying performance on ZH tasks. However, such framework-level gains remain constrained by the underlying backbone model. The performance gaps between GPT-5.5 and DeepSeek-V4-Pro indicate that modular orchestration cannot fully compensate for limitations in base model capability. Overall, while frameworks improve tool-use efficiency and workflow orchestration, final analytical performance is still bounded by the capacity of the core model.

\noindent\textbf{Skill Customization as Methodological Encapsulation.}
A key advantage of open-agentic workflows lies in customizable skill design, which contrasts with opaque proprietary pipelines. As shown in Figure~\ref{fig:task_taxonomy}, financial deep research tasks are highly heterogeneous, and different analytical tasks often require different combinations of tools. Configurable skill sets allow institutions to modularly embed domain rules and business expertise into data curation, analytical processing, and standardized report generation. This flexibility transforms general-purpose LLMs from generic conversational systems into more domain-adaptive specialized systems, making them better aligned with professional analysts' research workflows and reporting conventions.

\noindent\textbf{Architectural Evolution and Future Paradigms.}
The diagnostic splits and cross-lingual disparities shown in Figure~\ref{fig:score_comparison_butterfly_chart} point to a strategic divergence in enterprise Deep Research architectures. While closed-source proprietary products are effective at rapidly generating well-structured and professionally written reports in their dominant languages, high-stakes financial scenarios place stronger emphasis on traceability, verifiable logic, and localized data adaptation. Consequently, future enterprise-grade deep research may shift from fixed-pipeline monolithic systems toward highly configurable open-agentic pipelines. Such a paradigm can improve the factual reliability, professional presentation, and auditability of the final reports.

\subsection{The Illusion of Competence: Disentangling Reliability from Readability}
\noindent\textbf{The Paradox of Proprietary Models.} 
A critical finding from ICBCBench, visualized in Figure~\ref{fig:correlation_obj_sub}, is the stark discrepancy between Objective and Subjective performance. This divergence exposes a systemic ``illusion of competence'' within proprietary models: they excel at generating highly structured, authoritative narratives while simultaneously failing at rigorous factual extraction. For instance, Grok-3-deepsearch collapses to a mere 10.00 on EN Objective tasks, despite achieving Subjective scores exceeding 50.00. However, this decoupling simultaneously reveals their enduring strength. While open-agentic frameworks dominate verifiable data extraction, closed-source systems retain the absolute peak Subjective scores (e.g., OpenAI-o3-deep-research at 71.84 in EN, Gemini-deep-research at 65.69 in ZH). This suggests that internal generation pipelines, heavily optimized for long-context coherence and professional tone alignment, can aesthetically mask severe factual deficits.

\input{figures/correlation_obj_sub}\label{sec:paradox}

\noindent\textbf{The Insufficiency of Raw Factuality.}
Conversely, the data reveals that high objective accuracy does not inherently translate to high-quality subjective synthesis. MiroThinker ties for the highest EN objective score (52.50), yet its subjective score (53.15) significantly trails peers like OpenClaw and DeerFlow. This pattern highlights that raw factual extraction alone is insufficient for financial research. Producing an expert-level report demands narrative flow, structured argumentation, and domain-specific stylistic alignment, demonstrating that objective data extraction and subjective text synthesis represent fundamentally orthogonal dimensions of deep research intelligence.

\noindent\textbf{Implications for Financial Deep Research.}
The decoupling of these two capabilities underscores a critical cognitive bottleneck in current DR systems: a high subjective score guarantees readability but not reliability, while a high objective score ensures factual correctness but lacks communicative value. Consequently, advancing financial deep research requires moving beyond singular metric optimization. Recognizing this orthogonality paves the way for future methodologies to explicitly fuse deterministic, tool-driven verification with advanced narrative synthesis, effectively bridging the gap between objective reliability and subjective readability.

\section{Conclusion}
We introduce \textbf{ICBCBench}, an industry-aligned dual-track benchmark designed to rigorously evaluate financial Deep Research Agents. Our findings highlight a critical bifurcation in current AI systems: proprietary models excel at narrative synthesis but often suffer from an ``illusion of competence'' in factual extraction, whereas open-agentic frameworks demonstrate superior objective reasoning. By disentangling these orthogonal capabilities, we aim to catalyze the development of decoupled, hybrid architectures for the next generation of financial deep research systems.

\section*{Acknowledgments}
We are grateful to Hongsheng Gao, Deputy General Manager, and Chengyan Liu, Senior FinTech Expert, of the Software Development Center of Industrial and Commercial Bank of China, for their organizational support in facilitating this project.
We also thank Prof. David Lee Kuo Chuen, Professor at the Singapore University of Social Sciences, Founder of the Global FinTech Institute, and Chairman of the Board of Asia Pacific Exchange, for his valuable advice.
Li Guo acknowledges financial support from the National Natural Science Foundation of China (Project No.~72003040).

\medskip
\bibliographystyle{plainnat}
\bibliography{references}
\newpage

\appendix

\section{Dataset Details}
\label{sec:dataset_details}

\subsection{Difficulty Levels}
\label{subsec:appendix_difficulty}
\noindent\textbf{Difficulty Annotation.} We define four types of tools: \textit{Search} (retrieving information via search engines), \textit{Visit} (accessing and parsing web pages), \textit{Multi-modality} (processing image-based information using OCR or multimodal models), and \textit{Coding} (generating Python code for computation, visualization, or tool integration). Based on these capabilities, we categorize task difficulty into three levels: Level 1 (Easy), Level 2 (Medium), and Level 3 (Hard), with Level 3 accounting for over 70\% of the tasks, while Levels 1 and 2 together comprise no more than 30\%.
The difficulty of each task is determined by a combination of the number of information sources required, the number of tools involved, and the complexity of the reasoning process.

The criteria for each difficulty level is defined as follows:
\noindent\begin{itemize}[leftmargin=*, itemsep=4pt, topsep=0pt, parsep=0pt]
\item \textbf{Level 1 (Easy):} Typically involves 1–2 information sources, requires no tools or at most one tool, and can be solved within fewer than 5 steps.
\item \textbf{Level 2 (Medium):} Typically requires 3–5 information sources, may involve 2–3 tools, and can be solved within 5–8 steps.
\item \textbf{Level 3 (Hard):} Typically involves more than 5 information sources, may require multiple tools, and generally takes more than 8 steps to solve.
\end{itemize}

\input{tables/financial_taxonomy}
\input{tables/affiliation_distribution}
\input{tables/authoring_principles}
\input{tables/task_schema}
\input{tables/rating_scale}

\subsection{Objective Task Construction Details}
\label{subsec:objective_task_details}
To ensure that objective questions are grounded in real-world financial research needs and remain verifiable, we first collect over 20,000 financial research reports as the primary source materials and provide them to all task designers. To improve coordination and efficiency, we develop a dedicated platform to manage the entire construction process, including task submission, review, revision, and acceptance tracking. The full construction process consists of the following five stages.

\textbf{Stage 1: Initial Task Authoring.}
Tasks are constructed based on the principles in Table~\ref{tab:authoring_principles} by extracting key knowledge from financial reports and incorporating predefined factors such as tool usage, number of information sources, and reasoning complexity. The use of LLMs is encouraged to improve task quality. In addition to the task itself, designers are required to provide reasoning processes, source references, answer formats, tool annotations, and difficulty labels. The full task schema is shown in Appendix Table~\ref{tab:task_schema}.

\textbf{Stage 2: LLM-based Screening.}
We employ three SOTA models, namely Gemini-3-Pro-Preview~\cite{googledeepmind2025gemini31pro}, GPT-5.4~\cite{openai2025gpt54}, and Kimi-K2.5~\cite{team2026kimi}, for the first round of automated evaluation, aiming to filter out tasks that are overly simple or do not comply with the design principles. If a task can be directly solved by more than one LLM, the designer is required to increase its difficulty. Tasks that fail to meet the requirements after multiple rounds of refinement are discarded.

\textbf{Stage 3: Human Solving and Cross-review.}
Before formal human solving and review, all student annotators receive training and are provided with five high-quality annotated examples for reference. After passing the LLM-based screening in Stage 2, each task is independently solved and reviewed by at least three student annotators, who provide both answers and feedback on task quality. Following the practice of HLE, we further introduce a task recommendation scoring mechanism, where annotators assign a rating alongside their feedback, as detailed in Appendix Table~\ref{tab:rating_scale}.

\textbf{Stage 4: Task Refinement and Candidate Selection.}
Following the human solving and cross-review stage, each task receives at least three independent answers, recommendation scores, and feedback comments. 
Based on this feedback, task designers further refine the questions to improve clarity, verifiability, difficulty, and compliance with the authoring principles. 
Only tasks rated as \textit{High-quality} or above are retained as candidate tasks.

\textbf{Stage 5: Final Acceptance Review.}
After the first four stages, we obtain a candidate pool of objective tasks that have passed both LLM-based screening and human cross-review. 
The organizers then make the final acceptance decision, jointly considering LLM evaluations, human feedback, submitted answers, and recommendation scores. 
Tasks accepted at this stage are included in the final benchmark.

\subsection{Subjective Task Construction Details}
\label{subsec:subjective_task_details}
Enterprise dialogue queries often exhibit three common deficiencies: 
(1) \textbf{colloquial expression}, 
(2) \textbf{overly broad scope}, and 
(3) \textbf{lack of constraints}. 
Representative examples include:
\begin{itemize}[leftmargin=*, itemsep=2pt, topsep=2pt]
    \item \textbf{User A:} \textit{How can banks conduct digital operations in internet finance?}
    \item \textbf{User B:} \textit{How can healthcare insurance data support product development and pricing optimization in commercial insurance?}
    \item \textbf{User C:} \textit{Analyze financing challenges and solutions in PPP models based on asset relativity and comparative valuation theories.}
\end{itemize}

Such queries often require multiple rounds of interaction to clarify intent and produce usable research reports. To address these issues, we introduce a query refinement pipeline that maps raw queries to structured research prompts:
$
q_i' = \mathcal{R}(q_i),
$
where $\mathcal{R}(\cdot)$ denotes a query refinement operator instantiated using GPT-5.4 with curated exemplars. 
The refinement augments each query along three dimensions:
\begin{equation}
q' = q + c_{\text{context}} + c_{\text{constraints}} + c_{\text{structure}},
\end{equation}
where $c_{\text{context}}$ provides domain-specific background, $c_{\text{constraints}}$ introduces explicit analytical conditions, and $c_{\text{structure}}$ specifies the expected report format or output style. 
The detailed prompt is shown in Appendix Figure~\ref{fig:query_refinement_prompt}.

\section{Evaluation Details}
\label{sec:evaluation_details}

\subsection{Rubric Construction and LLM Judge Prompting}
\label{subsec:rubric_construction_details}

\textbf{Rubric construction.}
For each subjective task, we construct a task-specific rubric tailored to its report type and analytical objectives. Each rubric follows a 100-point scale and typically contains 4--6 high-level dimensions and 12--16 fine-grained sub-dimensions. The expert score $S_{\text{expert}}$ is obtained by aggregating scores across all rubric dimensions.

\textbf{Quality control and expert alignment.}
We adopt a multi-stage process to ensure rubric quality and alignment with expert judgment. Financial experts first draft initial rubrics based on task descriptions, analytical objectives, and representative report samples. GPT-5.4 is then used to refine rubric structure, improve granularity, and clarify scoring criteria. The refined rubrics are reviewed by at least three additional domain experts, and the final versions are consolidated by the organizers. Participating institutions are listed in Table~\ref{tab:affiliation_distribution}, and an example rubric is shown in Appendix Figure~\ref{fig:appendix_case_expert_rubric}.

\textbf{LLM-as-a-Judge with rubric grounding.}
We use Gemini-3.1-Pro-Preview as the evaluator for fine-grained quantitative scoring. 
The evaluator is instructed to score each report strictly according to the predefined rubric sub-dimensions and to provide a justification grounded in the report content for each score. This rubric-grounded prompting aims to reduce bias, inconsistency, and hallucination in LLM-based evaluation. The detailed judge prompt is shown in Appendix Figure~\ref{fig:subjective_judge_prompt}.

\subsection{Source Authority and Timeliness Scoring}
\label{subsec:source_scoring}

We compute source quality from two dimensions, authority and timeliness, and combine them into a unified score.

\textbf{Source authority.} We categorize cited URLs into four provenance-based tiers. Tier 1 includes highly authoritative sources, such as official institutions and leading financial outlets. Tiers 2 and 3 include reputable institutional sources and general media sources, respectively, while Tier 4 covers all remaining sources. Each tier is assigned a normalized authority score $S_{\text{auth}} \in [0,1]$.

\textbf{Information timeliness.}
To account for the time sensitivity of financial information, we model source recency using an exponential decay function:
\begin{equation}
S_{\text{time}} = e^{-\alpha \cdot \Delta t},
\end{equation}
where $\Delta t$ denotes the source age in days and $\alpha$ controls the decay rate. 
We set $\alpha = 0.002$, corresponding to a half-life of approximately one year.

For each task $t$, let $m_t$ denote the number of successfully scraped URLs. 
The per-task source quality score is computed as:
\begin{equation}
S_{\text{source}}^{(t)} =
\begin{cases}
\frac{1}{m_t} \sum_{i=1}^{m_t} S_{\text{auth}}^{(i)} \cdot S_{\text{time}}^{(i)}, & m_t > 0 \\
0, & m_t = 0
\end{cases}
\end{equation}
where $S_{\text{auth}}^{(i)}$ and $S_{\text{time}}^{(i)}$ denote the authority and timeliness scores of the $i$-th URL, respectively. 
The overall source quality score is:
\begin{equation}
S_{\text{source}} = \frac{100}{T} \sum_{t=1}^{T} S_{\text{source}}^{(t)},
\end{equation}
where $T$ is the number of subjective tasks, and the factor 100 scales the score to $[0,100]$.

\subsection{Evaluation Model List}
\label{subsec:appendix_model_list}
\input{tables/model_versions}
The closed-source models include Gemini-deep-research~\cite{google2024geminideepresearch}, o3-deep-research~\cite{openai2025o3deepresearch}, Perplexity-deep-research~\cite{perplexity2025deepresearch}, Grok-3-deepsearch~\cite{xai2025grok3}, Doubao-deep-research~\cite{doubao_chat_2026}, Qwen-deep-research~\cite{qwen2025deepresearch}, Kimi-deep-research~\cite{moonshot2025kimi_researcher}, as well as advanced general-purpose models such as Gemini-3-pro-preview~\cite{google2025gemini3}, GPT-5.4~\cite{openai2025gpt54}, GPT-5.5~\cite{openai2026gpt55},  Claude-opus-4-7~\cite{anthropic2026opus47}, and Kimi-k2.5~\cite{team2026kimi}. 

The open-source and framework-based systems include Jina-deepsearch~\cite{jina2025deepsearch}, Tongyi-deepresearch-30b-a3b~\cite{team2025tongyi}, MiroThinker~\cite{team2025mirothinker}, DeepSeek-V3.2~\cite{liu2025deepseek}, and DeepSeek-V4-Pro~\cite{deepseekai2026deepseekv4}. 
We also evaluate open agentic frameworks, including DeerFlow~\cite{bytedance2026deerflow} and OpenClaw~\cite{openclaw2026openclaw}, instantiated with GPT-5.5~\cite{openai2026gpt55} and DeepSeek-V4-Pro~\cite{deepseekai2026deepseekv4} backbones. Detailed model and framework configurations, together with the release dates used in our experiments, are summarized in Table~\ref{tab:model_versions}.

\subsection{Evaluation Metrics of Human Consistency}
\label{subsec:append_evaluation_detail}
The details of the evaluation metrics of human consistency are as follows.
\begin{itemize}
    \item \textbf{Spearman's Rank Correlation Coefficient ($\rho$):} We use this to measure the relative ranking consistency. Unlike Pearson correlation, which assumes linearity and can be skewed by absolute scaling differences, Spearman strictly evaluates whether the LLM correctly preserves the ordinal ranking of the reports.
    \item \textbf{Mean Absolute Error (MAE):} To complement the relative ranking, MAE is employed to quantify the systemic deviation in absolute scores, directly reflecting how closely the LLM's scoring scale aligns with rigorous human standards.
    \item \textbf{Pairwise Agreement Rate:} We also report the pairwise win/tie/loss agreement, which measures how often the LLM's binary preference between any pair of reports matches the consensus of human experts, providing an intuitive gauge of decision reliability. 
\end{itemize}

\subsection{Open-Agentic Framework Configurations and Execution Issues}
\label{subsec:framework_configurations}
We evaluate MiroThinker, DeerFlow, and OpenClaw under controlled local deployment settings. 
Table~\ref{tab:agent_skills} lists the complete set of agent skills used in DeerFlow and OpenClaw.

\input{tables/agent_skills}

\paragraph{Framework configurations.}
\begin{itemize}[leftmargin=*, itemsep=2pt, topsep=2pt]
    \item \textbf{MiroThinker.} 
    We use the official framework with the MiroThinker-1.7-235B model deployed locally, Serper for search, and Jina for web crawling. 
    The official evaluation script is adapted to run ICBCBench tasks.

    \item \textbf{DeerFlow.} 
    We deploy the official DeerFlow framework locally, using Tavily for search and Jina for web crawling. 
    The summarization threshold is set to 150,000 tokens, and the tool invocation limit is increased to 50 calls to reduce context compression and premature termination. 
    Each task is evaluated in a separate conversation to avoid cross-question interference.

    \item \textbf{OpenClaw.} 
    We evaluate OpenClaw v2026.4.8 with Tavily as the search tool. 
    DeerFlow's deep-research skill is incorporated into OpenClaw to improve report standardization, readability, and traceability. 
    The agent workspace is reset before each task to prevent memory interference.
\end{itemize}

\paragraph{Execution issues.}
During report generation with closed-source Deep Research APIs and locally deployed frameworks, we observed several execution issues that affected stability and reproducibility.

\begin{itemize}[leftmargin=*, itemsep=2pt, topsep=2pt]
    \item \textbf{API errors.}
    High-concurrency API calls occasionally led to failed or incomplete responses, requiring retries or manual filtering during data collection.

    \item \textbf{Report generation failures.}
    Some queries failed to produce coherent final reports. 
    In a few cases, models such as Kimi-k2.5-thinking returned intermediate tool-calling traces or planning prompts rather than final answers.

    \item \textbf{MiroThinker.}
    Generated outputs occasionally used inconsistent formatting, alternating between Markdown and LaTeX-style structures without a unified output standard.

    \item \textbf{DeerFlow.}
    DeerFlow showed several execution instability issues, including occasional ordering inconsistencies when merging multiple Markdown files, frequent tool invocations that could exceed preset limits, and report-style outputs for objective question-answering tasks. 
    Some runs also included redundant or low-utility tool calls, which could lead to early termination.

    \item \textbf{OpenClaw.}
    OpenClaw was generally more stable in tool usage, with around 10 tool invocations per task, but occasionally suffered from long response times. 
    We therefore extended the timeout limit to 1800 seconds to allow complete responses.
\end{itemize}

These observations highlight practical challenges in evaluating Deep Research systems, where performance depends not only on model capability but also on orchestration stability, tool-use efficiency, and output-format consistency.

\input{figures/score_comparison_heatmap}
\input{tables/results_subset_en}
\input{tables/results_subset_zh}
\input{tables/results_private_set}
\section{Extended Results and Analysis}
\label{sec:extended_results_and_analysis}
This section presents granular evaluation results for ICBCBench. 
Figure~\ref{fig:score_comparison_heatmap} provides a diagnostic heatmap highlighting the systemic disparity between Objective reasoning and Subjective generation. Tables~\ref{tab:results_subset_en} and \ref{tab:results_subset_zh} detail performance breakdowns for the English and Chinese subsets across fine-grained dimensions. Finally, to validate generalization and rule out overfitting, Table~\ref{tab:results_private_set} reports performance on our strictly sequestered private hold-out set.

\subsection{Granular Metric Insights}
\label{subsec:granular_insights}
\paragraph{Subjective Performance: Citation and Source Quality.} 
The narrative strength of proprietary models is largely driven by their exceptional \textit{Citation Consistency}. For instance, OpenAI-o3-deep-research dominates this metric in both EN (74.47) and ZH (79.35) subsets, indicating superior mechanisms for grounding synthesized text. Interestingly, while Jina-deepsearch underperforms overall, it achieves the highest \textit{Source Quality} scores (27.54 in EN, 31.83 in ZH), suggesting an effective initial retrieval strategy bottlenecked by its subsequent synthesis capabilities.

\paragraph{Objective Bottlenecks: Text-Only vs.\ Multimodal Queries.}
Objective scores reveal a consistent performance drop when moving from \textit{Text-Only} queries to the \textit{All} category, which additionally includes multimodal questions involving images, charts, tables, and visually presented financial information. 
For example, Gemini-deep-research drops from 61.76 to 52.50 in the ZH track, and DeerFlow(+GPT-5.5) declines from 57.14 to 52.50 in the EN track. This degradation highlights multimodal financial reasoning as a key bottleneck, requiring models to extract precise visual evidence and integrate it with textual reasoning.

\subsection{Generalization on Private Hold-out Set}
\label{subsec:private_set_results}
Evaluation on the unreleased private hold-out set (Table~\ref{tab:results_private_set}) broadly corroborates the main findings while revealing non-trivial differences from the public benchmark. Open-agentic frameworks remain highly competitive, particularly on objective tasks, with OpenClaw(+DeepSeek-V4-Pro) achieving the highest Overall score in the Global (EN) scenario and OpenClaw(+GPT-5.5) leading in the Chinese (ZH) scenario. 
Among closed-source systems, Gemini-deep-research remains the most robust overall.

However, private-set scores do not perfectly mirror public-set performance. 
Several models exhibit changes in absolute scores and rankings across languages and task types, reflecting differences in task composition, difficulty, and the smaller hidden split. 
These discrepancies indicate that the private set provides a complementary stress test rather than a direct replication of the public benchmark, helping assess generalization and reduce the risk of benchmark overfitting.

\input{figures/correlation_accuracy_vs_rmsce_public}
\subsection{Calibration Error}
\label{subsec:calibration_error}
\paragraph{Accuracy vs Calibration Quality.}
Figure~\ref{fig:correlation_accuracy_vs_rmsce_public} visualizes the relationship between accuracy and calibration error on the objective subset across both global and Chinese subsets. Only six systems fall within the ideal zone, demonstrating not only strong factual reasoning but also well-calibrated confidence, a critical property for deploying LLMs in high-stakes financial decision-making. The top three positions in Figure~\ref{fig:correlation_accuracy_vs_rmsce_public} are occupied by OpenClaw paradigm systems and followed by Gemini-deep-research.

\input{tables/objective_public_CalibErr}
\paragraph{Calibration Deficiencies.}
The majority of models cluster in the bottom-right of the plot, exhibiting poor calibration. Deep Research products such as Grok-3-deepsearch, Qwen-deep-research, and Tongyi-deepresearch-30b-a3b suffer from extreme calibration errors exceeding $80\%$, indicating that their confidence scores are essentially uncorrelated with correctness. Even some top-tier systems including DeerFlow(+DeepSeek-V4-Pro) and Kimi-deep-research fall outside the ideal zone due to miscalibrated confidence, despite achieving moderate accuracy. The full numerical results are provided in Appendix Table~\ref{tab:objective_public_CalibErr}.

\input{tables/results_domain_specific}
\subsection{Domain-Specific Performance}
\label{subsec:domain_specific_performance}
Table~\ref{tab:results_domain_specific} presents the domain-specific performance of evaluated systems across major financial sectors, including Banking, Securities, Insurance, and Other Financial Services. These results provide additional insights into how different systems generalize across heterogeneous financial domains with varying levels of reasoning complexity, domain knowledge requirements, and factual grounding challenges.


\section{Related Work}
\label{related_work}
\paragraph{General Deep Research Benchmarks.}
Recent benchmarks evaluate deep research capabilities using short, verifiable closed-ended questions. HLE~\cite{phan2025humanity}, GAIA~\cite{mialon2023gaia}, and BrowseComp~\cite{Wei2025BrowseCompAS} emphasize answer-based evaluation with well-defined ground truth, enabling scalable and objective measurement. However, they mainly capture factual correctness and basic reasoning, and are limited in assessing complex analysis and long-form report generation required in real-world scenarios.

\input{tables/benchmark_comparison}

\paragraph{Long-form Report Evaluation and LLM-as-a-Judge.}
To better reflect real-world research tasks, recent work has shifted toward evaluating long-form report generation. Benchmarks such as DeepResearch Bench~\cite{Du2025DeepResearchBA}, DR.BENCH~\cite{Yao2025DrBA}, LiveResearchBench~\cite{wang2025liveresearchbenchlivebenchmarkusercentric}, and DRACO~\cite{Zhong2026DRACOAC} adopt report-based evaluation with citation-aware metrics and rubric-based assessment. Extensions to multimodal settings have also emerged, including MMDeepResearch-Bench~\cite{huang2026mmdeepresearchbenchbenchmarkmultimodaldeep} and Vision-DeepResearch Benchmark~\cite{zeng2026visiondeepresearchbenchmarkrethinkingvisual}. However, these approaches either rely on fixed evaluation dimensions or lack flexible, expert-aligned frameworks, limiting their ability to support structured, domain-specific report analysis.

\paragraph{Financial Domain Benchmarks.}
Several recent works attempt to introduce financial-specific evaluation settings, including FinRpt~\cite{Jin2025FinRptDE} and FinResearchBench~\cite{Sun2025FinResearchBenchAL}. These benchmarks incorporate financial scenarios and report-generation tasks, but remain limited in scope and evaluation methodology. FinRpt focuses primarily on equity research reports with relatively simple evaluation protocols, while FinResearchBench adopts a logic tree-based agent evaluation framework that lacks broad validation from domain experts. More generally, existing benchmarks in finance either rely on a single evaluation paradigm or fail to capture the full complexity of real-world financial research workflows.

\paragraph{Our Contribution.}
In contrast to prior work, ICBCBench introduces a unified dual-track evaluation paradigm that integrates objective question answering with subjective report generation. Furthermore, we propose a hybrid evaluation framework for long-form reports that combines expert-defined rubrics, citation consistency checking, and source quality verification. By grounding both task design and evaluation in real-world financial practice and domain expert knowledge, ICBCBench provides a comprehensive and industry-aligned benchmark for financial deep research.

\section{Limitations, and Future Work}
\label{limitations_and_future_work}
Despite its rigorous design, this study faces limitations regarding the temporal degradation of financial data, the computational overhead of multi-agent workflows, and the inherent difficulty of computationally evaluating contrarian market insights. To address these challenges, future work will transition towards \textit{live benchmarking} environments to evaluate DRAs against real-time market dynamics. Furthermore, we advocate for the development of \textit{hybrid architectures} that fuse the deterministic fact-checking of open-agentic frameworks with the sophisticated long-context synthesis of frontier models, paving the way for truly autonomous financial research.

\section{Case Studies}
\label{sec:case_studies}
This section presents representative examples from ICBCBench to illustrate our task diversity and evaluation rigor. Figure~\ref{fig:appendix_case_objective} shows two objective tasks requiring precise, verifiable financial reasoning. Figure~\ref{fig:appendix_case_subjective} presents two subjective report-generation tasks derived from real industry needs, spanning banking digital operations and AI-driven industry transformation. To illustrate the evaluation design, Figure~\ref{fig:appendix_case_expert_rubric} provides the expert rubric for the banking digital operations task, demonstrating how ICBCBench assesses analytical depth, practical relevance, factual grounding, and structured reporting quality through task-specific criteria.
\vspace{1em}
\input{prompts/appendix_case_objective}
\newpage
\input{prompts/appendix_case_subjective}
\newpage
\input{prompts/appendix_case_expert_rubric}

\section{Prompts}
\label{sec:prompts}
This section presents the complete set of prompts utilized throughout the ICBCBench framework. To ensure full experimental reproducibility, we provide the exact system-level instructions used for agentic query refinement (Figure~\ref{fig:query_refinement_prompt}) together with the automated evaluation prompts for both objective and subjective judging (Figures~\ref{fig:judge_prompt} and \ref{fig:subjective_judge_prompt}).
\input{prompts/query_refinement_prompt}
\input{prompts/objective_evaluation_prompt}
\input{prompts/subjective_evaluation_prompt}

%% file: figures/main_figure.tex
\begin{figure}[t]
  \centering
  \includegraphics[width=\linewidth, trim=50 100 50 80, clip]{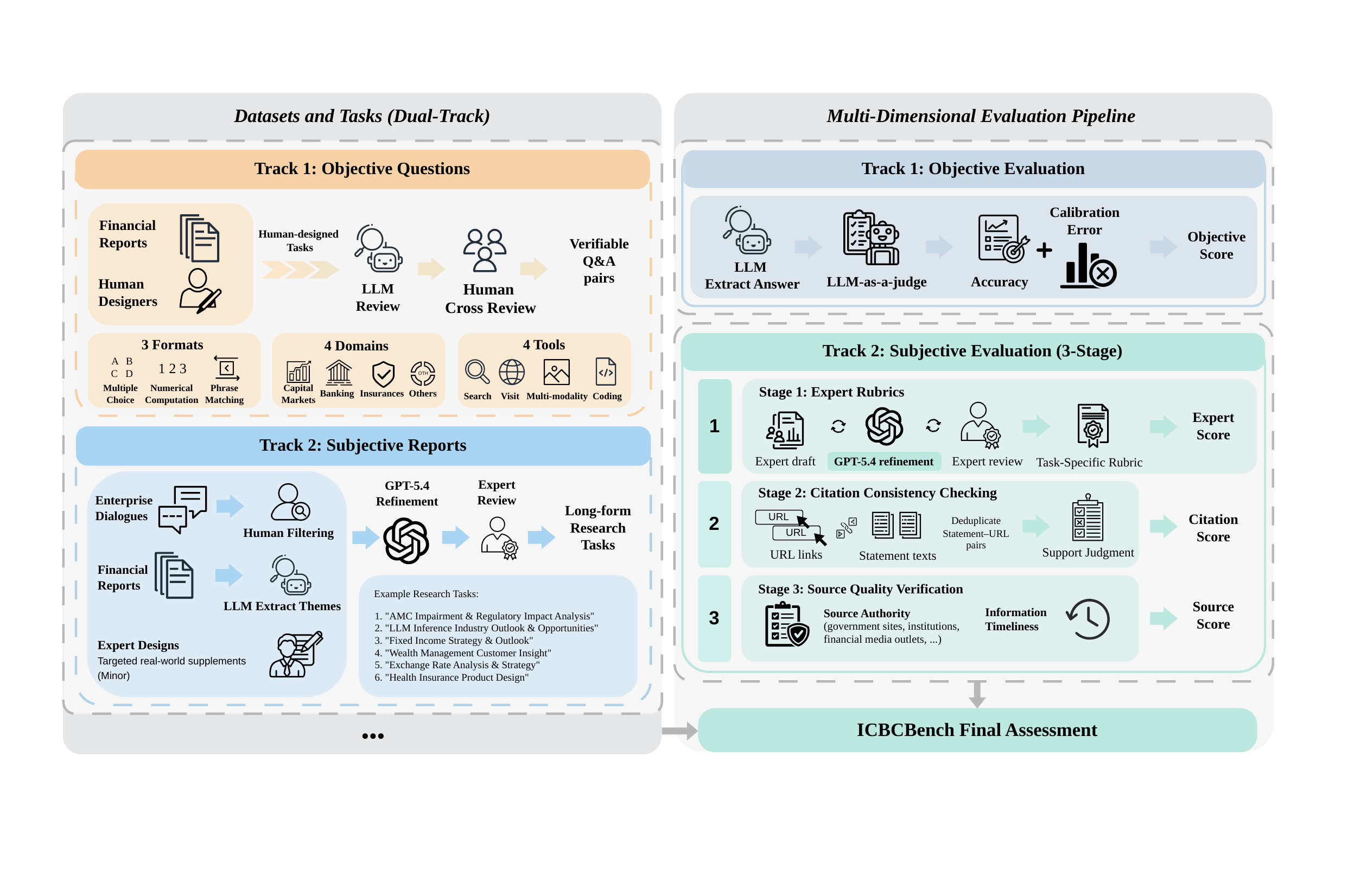} 
  \caption{\textbf{Overview of the ICBCBench construction and evaluation pipeline.}
  ICBCBench is built through industry--academia collaboration, integrating enterprise dialogues, financial reports, and expert-designed tasks into a dual-track benchmark. 
  Objective questions with verifiable answers are evaluated for factual correctness and confidence calibration, while subjective reports are assessed through an expert-aligned framework combining task-specific rubrics, citation consistency checking, and source quality verification.}
  \label{fig:icbcbench_pipeline}
  \vspace{-0.3cm}
\end{figure}

%% file: figures/task_taxonomy.tex
\begin{figure*}[t!]
  \centering
  
  \begin{subfigure}[b]{0.38\linewidth}
    \centering
    \includegraphics[width=\linewidth, trim=2cm 2cm 2cm 2cm, clip]{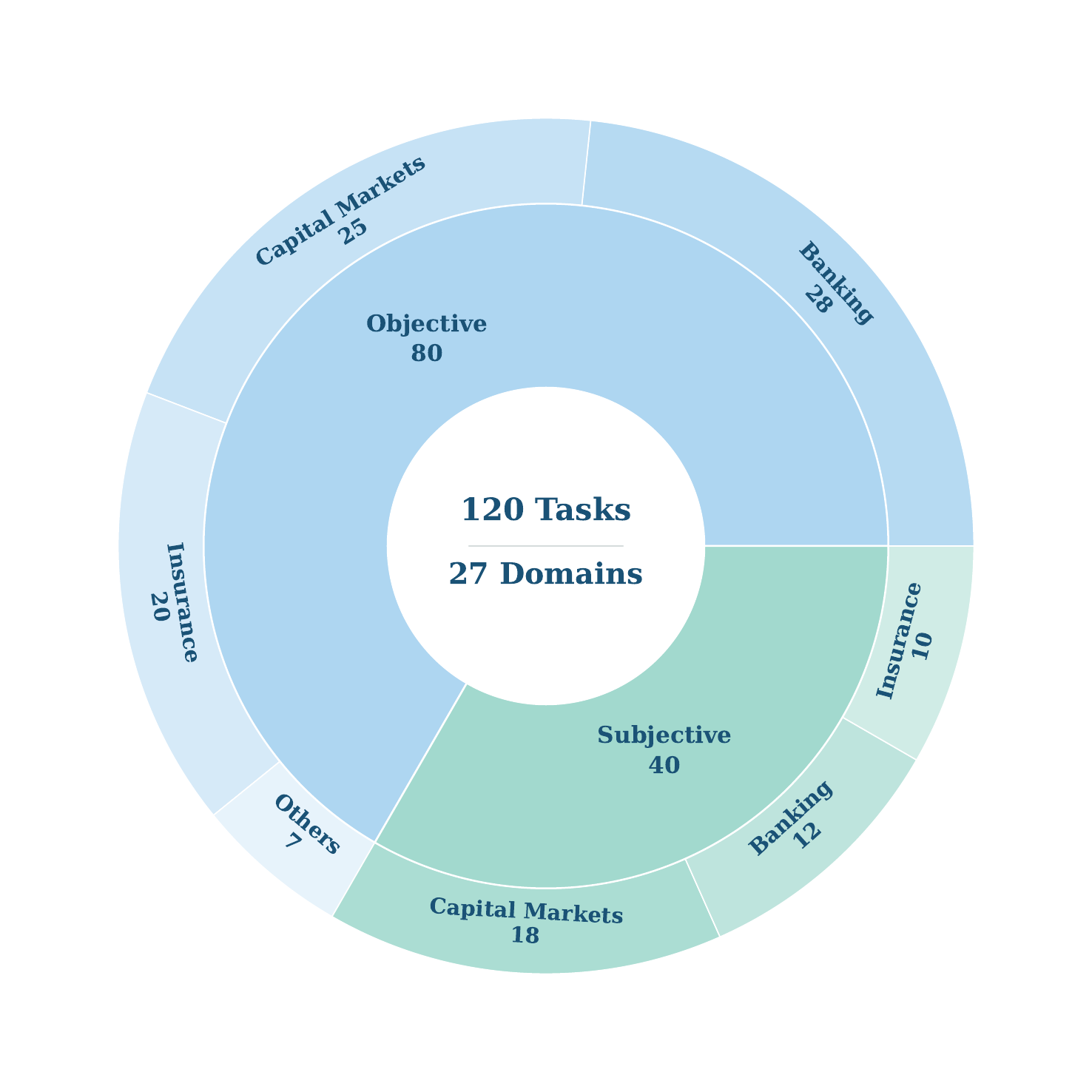}
    \vspace{12pt}
    \caption{Macro Task Distribution}
    \label{fig:task_a_macro}
  \end{subfigure}
  \hfill
  \begin{subfigure}[b]{0.58\linewidth}
    \centering
    \includegraphics[width=\linewidth, trim=0cm 0cm 4cm 0cm, clip]{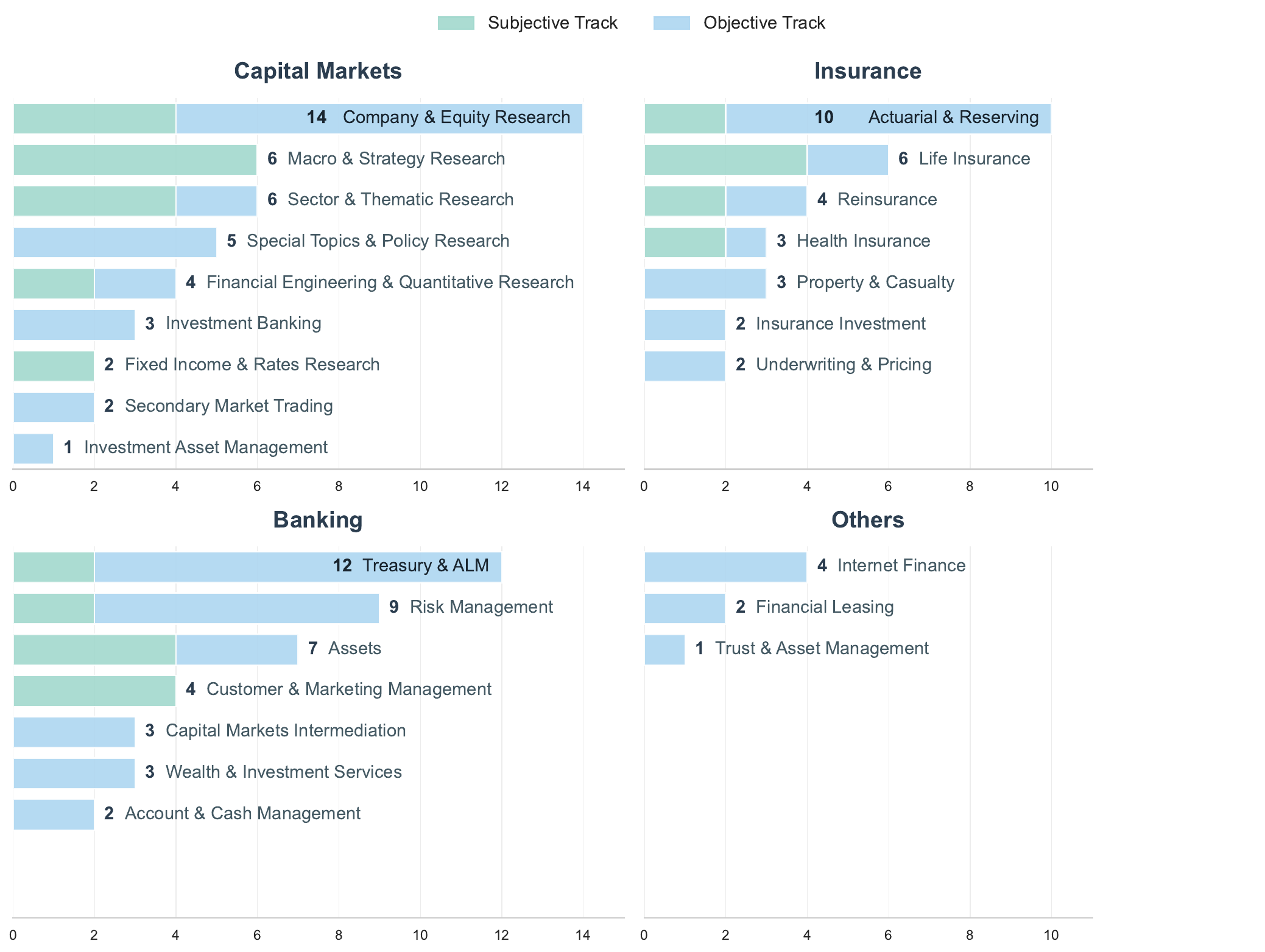}
    \vspace{-12pt}
    \caption{Sub-Domain Taxonomy}
    \label{fig:task_b_granular}
  \end{subfigure}
  
  \caption{\textbf{Multi-dimensional taxonomy of the public ICBCBench tasks.} (a) Macro distribution of the \textbf{120 public tasks} across Objective and Subjective tracks and four primary financial domains. (b) Granular task allocation across the \textbf{27 sub-domains} covered by the public subset. The complete taxonomy is provided in Appendix Table~\ref{tab:financial_taxonomy}.}
  \label{fig:task_taxonomy}
\vspace{-0.4cm}
\end{figure*}

%% file: tables/main_results.tex
\begin{table*}[t]
\centering
\small
\setlength{\tabcolsep}{5pt}
\renewcommand{\arraystretch}{1.08}

\caption{\textbf{Main results on ICBCBench.} We report performance on global (EN) and Chinese (ZH) market scenarios. Objective evaluates closed-ended tasks, while Subjective assesses open-ended financial research reports. The best and second-best scores are highlighted in \textbf{bold} and \underline{underline}, respectively. Higher is better for all metrics.}
\label{tab:main_results}

\begin{tabular}{lcccccc}
\toprule
\multirow{2}{*}{\textbf{System}} & \multicolumn{3}{c}{\textbf{Global (EN)}} & \multicolumn{3}{c}{\textbf{Chinese (ZH)}} \\
\cmidrule(lr){2-4} \cmidrule(lr){5-7}
& \textbf{Objective} & \textbf{Subjective} & \textbf{Overall} & \textbf{Objective} & \textbf{Subjective} & \textbf{Overall} \\
\midrule
\rowcolor{gray!20}
\multicolumn{7}{c}{\textbf{\textit{Closed}}} \\
Gemini-deep-research                     &           50.00 &           64.77 & \underline{57.38} &           52.50 &  \textbf{65.69} & \underline{59.09} \\
OpenAI-o3-deep-research                  &           37.50 &  \textbf{71.84} &           54.67 &           32.50 & \underline{63.12} &           47.81 \\
Kimi-deep-research                       &           35.00 &           60.19 &           47.59 &           35.00 &           54.44 &           44.72 \\
Doubao-deep-research                     &           37.50 &           52.93 &           45.22 &           20.00 &           52.61 &           36.30 \\
GPT-5.5                                  &           27.50 &           62.69 &           45.09 &           27.50 &           57.33 &           42.41 \\
Claude-opus-4-7                          &           25.00 &           63.71 &           44.36 &           20.00 &           60.83 &           40.41 \\
Perplexity-deep-research                 &           22.50 &           63.17 &           42.84 &           22.50 &           48.85 &           35.67 \\
Gemini-3.1-pro-preview                   &           22.50 &           59.53 &           41.02 &           12.50 &           58.25 &           35.38 \\
Grok-3-deepsearch                        &           10.00 &           56.43 &           33.22 &            5.00 &           50.40 &           27.70 \\
Qwen-deep-research                       &            2.50 &           51.59 &           27.05 &           17.50 &           48.25 &           32.88 \\
\midrule
\rowcolor{gray!20}
\multicolumn{7}{c}{\textbf{\textit{Open}}} \\
DeerFlow(+GPT-5.5)                       &  \textbf{52.50} &           64.85 &  \textbf{58.67} & \underline{60.00} &           57.67 &           58.84 \\
OpenClaw(+GPT-5.5)                       &           50.00 &           59.60 &           54.80 &  \textbf{67.50} &           59.25 &  \textbf{63.38} \\
MiroThinker                              &  \textbf{52.50} &           53.15 &           52.83 &           45.00 &           43.88 &           44.44 \\
OpenClaw(+DeepSeek-V4-Pro)               &           37.50 & \underline{65.79} &           51.65 &           57.50 &           57.36 &           57.43 \\
DeerFlow(+DeepSeek-V4-Pro)               &           27.50 &           65.71 &           46.60 &           55.00 &           58.08 &           56.54 \\
Jina-deepsearch                          &           37.50 &           47.51 &           42.50 &           35.00 &           50.89 &           42.95 \\
Kimi-k2.5                                &           17.50 &           64.81 &           41.16 &           10.00 &           61.60 &           35.80 \\
DeepSeek-V4-Pro                          &            5.00 &           49.09 &           27.05 &           15.00 &           55.59 &           35.30 \\
Tongyi-deepresearch-30b-a3b              &            2.50 &           46.69 &           24.59 &            5.00 &           42.16 &           23.58 \\
\bottomrule
\end{tabular}
\end{table*}

%% file: figures/score_comparison_butterfly_chart.tex
\begin{figure*}[t]
  \centering
  \includegraphics[width=0.95\linewidth, trim=5 5 5 5, clip]{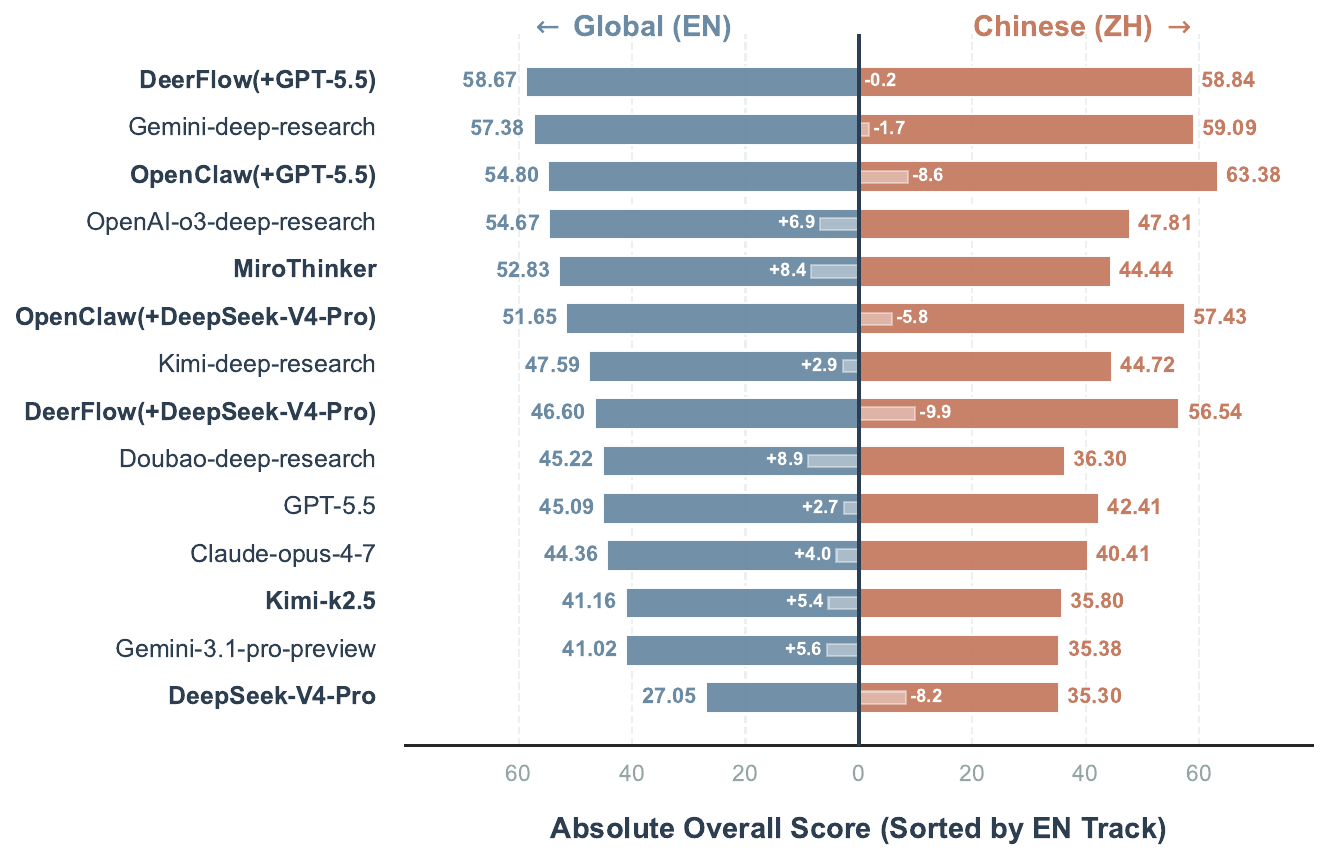} 
\caption{\textbf{Symmetric overall performance comparison and cross-lingual localization gap on ICBCBench.} 
The blue (left) and red (right) bars represent absolute overall scores on the Global (EN) and Chinese (ZH) tracks, respectively, with models ranked by EN performance. 
The overlaying white bars quantify the localization bias ($\Delta = \text{EN} - \text{ZH}$), where a positive value indicates English-centric dominance and a negative value reflects Chinese-first optimization. Bold model names denote open-source frameworks.}
  \vspace{-0.3cm}
  \label{fig:score_comparison_butterfly_chart}
\end{figure*}

%% file: human_consistency/human_consistency.tex
\begin{figure*}[htb]
\centering

\begin{subfigure}[b]{0.48\textwidth}
    \centering
    \includegraphics[width=\linewidth, trim=0cm 0cm 1cm 0cm, clip]{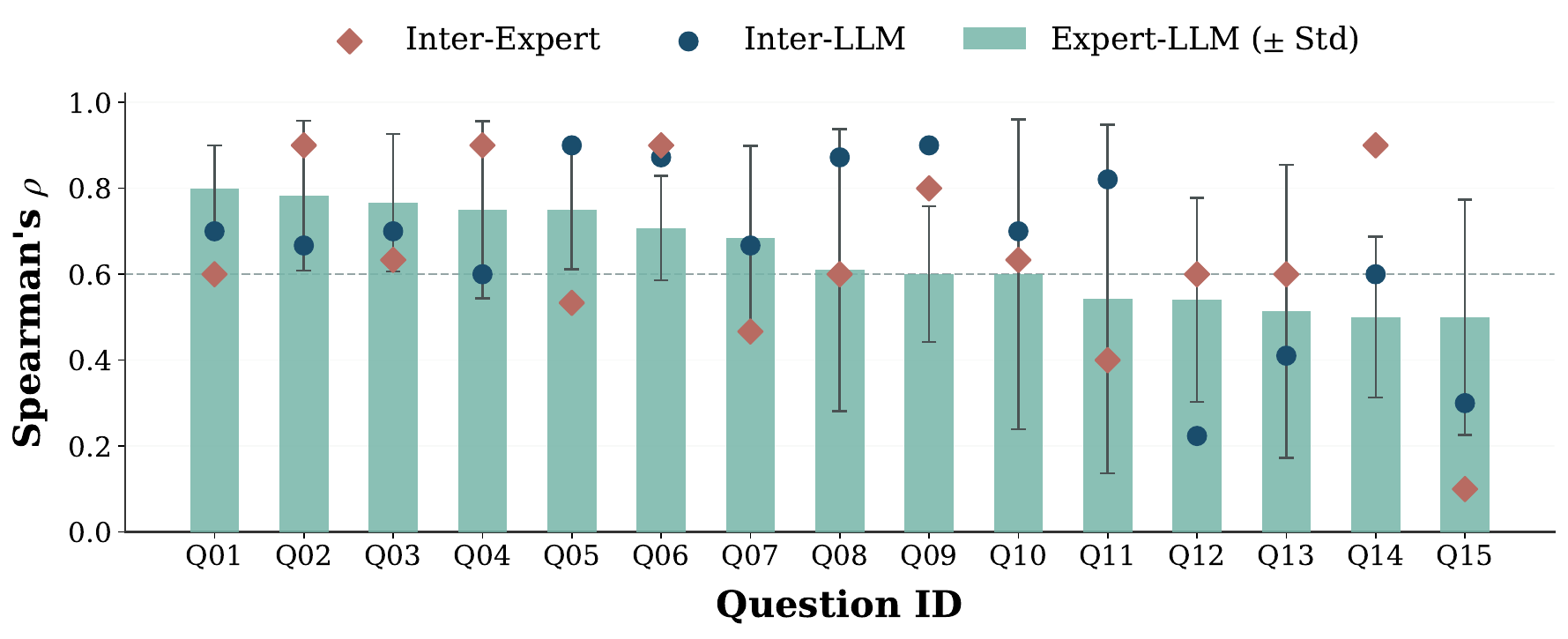}
    \caption{Rank Correlation (Spearman's $\rho$)}
    \label{fig:human_consistency_rank}
\end{subfigure}
\hfill
\begin{subfigure}[b]{0.48\textwidth}
    \centering
    \includegraphics[width=\linewidth, trim=0cm 0cm 1cm 0cm, clip]{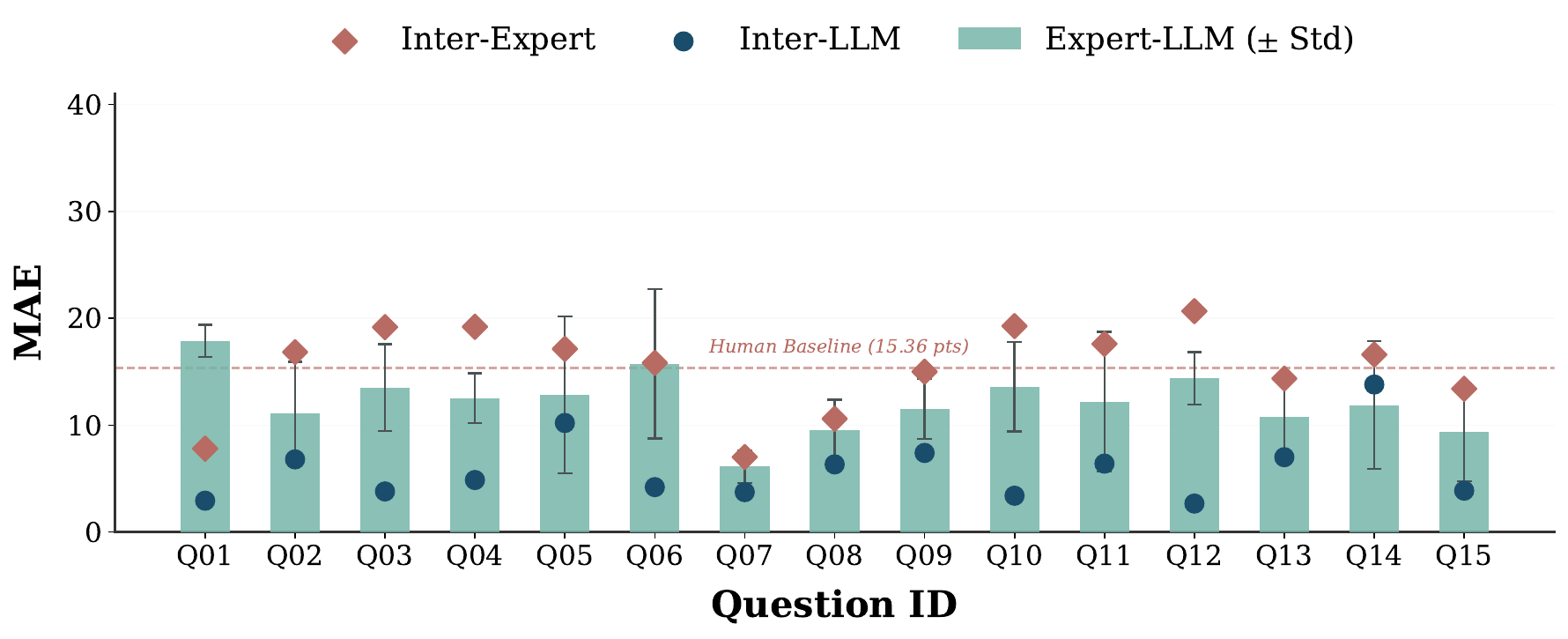}
    \caption{Absolute Score Deviation (MAE)}
    \label{fig:human_consistency_score}
\end{subfigure}

\caption{\textbf{Multi-dimensional human consistency analysis of LLM judges in 15 randomly selected samples from ICBCBench.} 
(a) Relative rank correlation using Spearman's $\rho$, showing high ranking alignment between human experts and LLMs. 
(b) Systemic deviation in absolute scores evaluated by Mean Absolute Error (MAE), indicating that the Expert-LLM score deviation falls strictly within the natural variance of the human baseline (inter-expert deviation).}
\label{fig:human_consistency_analysis}
\end{figure*}

%% file: human_consistency/overall_consistency.tex
\begin{table}[t]
\centering
\caption{\textbf{Overall Consistency Metrics on ICBCBench.} We report macro-level evaluation across ranking correlation (Spearman's $\rho \uparrow$), pairwise agreement (PAR $\uparrow$), and score deviation (MAE $\downarrow$). Results highlight that \textbf{Expert-LLM} alignment ($\rho=0.643$, PAR=0.729) successfully reaches the \textbf{Inter-Expert} consensus ceiling. 
}
\label{tab:overall_consistency}
\small
\setlength{\tabcolsep}{8pt}
\renewcommand{\arraystretch}{1.2}
\begin{tabular}{lccc}
\toprule
\textbf{Evaluation Dimension} & \textbf{Spearman's $\rho$ ($\uparrow$)} & \textbf{Agreement PAR ($\uparrow$)} & \textbf{Score MAE (pts) ($\downarrow$)} \\ 
\midrule
Inter-Expert  & 0.638 & 0.711 & 15.36 \\ 
Inter-LLM     & 0.662 & 0.733 & \phantom{0}5.83 \\ 
\midrule
\textbf{Expert-LLM (Alignment)} & \textbf{0.643} ($\pm$ 0.227) & \textbf{0.729} ($\pm$ 0.120) & \textbf{12.20} ($\pm$ 4.09) \\ 
\bottomrule
\end{tabular}
\end{table}

%% file: figures/correlation_obj_sub.tex
\begin{figure*}[htb]
\centering
\includegraphics[width=0.96\textwidth, trim=0cm 0cm 0cm 0cm, clip]{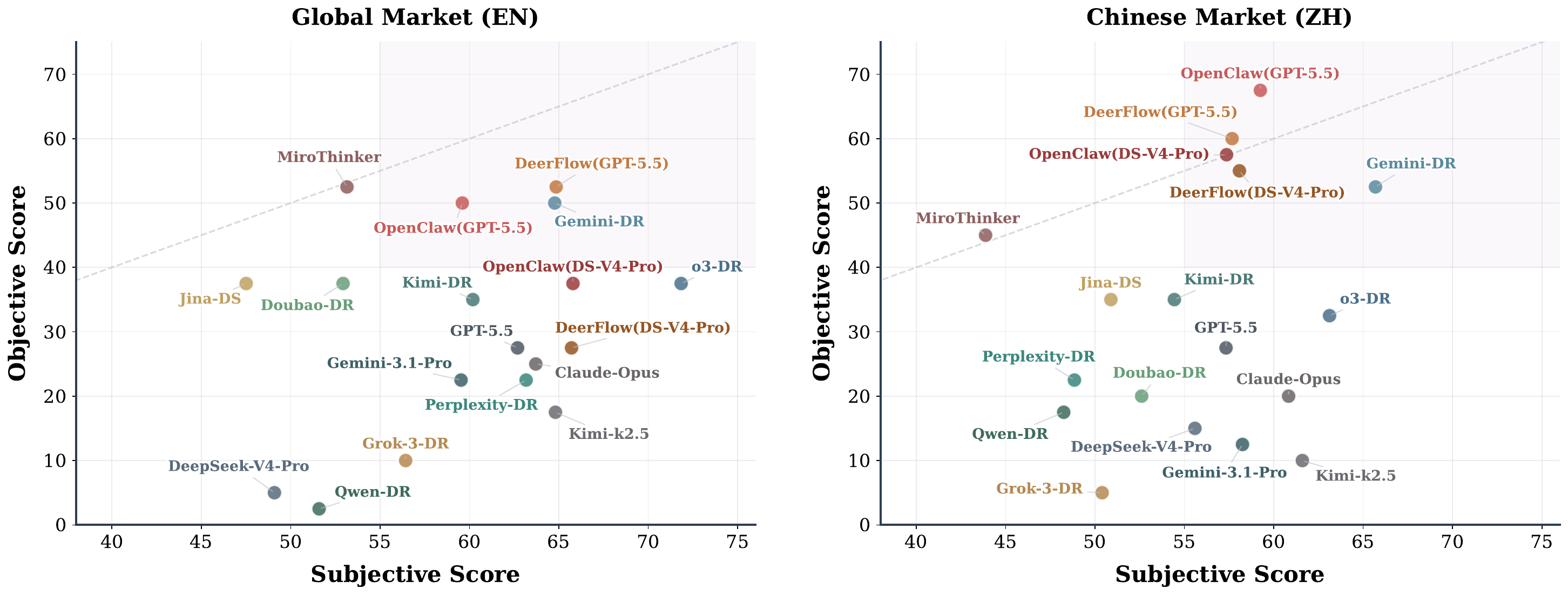}
\caption{\textbf{Correlation between Objective and Subjective Evaluation Tracks.} The scatter plots illustrate the alignment between models' objective scores and subjective evaluations across English and Chinese tasks. The strong positive correlation demonstrates the systematic reliability and robust bilingual evaluation capabilities of the ICBCBench framework.}
\label{fig:correlation_obj_sub}
\end{figure*}

%% file: tables/financial_taxonomy.tex
\begin{table}[t]
\centering
\small
\setlength{\tabcolsep}{4pt}
\renewcommand{\arraystretch}{1.1}

\caption{Taxonomy of financial research domains in ICBCBench.}
\label{tab:financial_taxonomy}

\begin{tabularx}{\linewidth}{lX}
\toprule

\textbf{Domains} & \textbf{Subdomains} \\

\midrule

\textbf{Capital Markets} 
& Primary Market, Secondary Trading, Asset Management, Investment Banking, Custody \& Clearing, Macro \& Strategy, Sector \& Thematic, Equity Research, Fixed Income \& Rates, Quantitative \& Financial Engineering, Policy \& ESG Research, Client \& Product Research \\

\textbf{Banking} 
& Liabilities, Assets, Payments \& Settlement, Account \& Cash Management, Wealth \& Investment Services, Capital Markets Intermediation, Treasury \& ALM, Customer \& Marketing Management, Risk Management \\

\textbf{Insurance} 
& Life Insurance, Health Insurance, Property \& Casualty, Reinsurance, Underwriting \& Pricing, Claims \& Fraud, Actuarial \& Reserving, Insurance Investment \\

\textbf{Other Financial Services} 
& FinTech, Inclusive Finance, Credit Guarantee, Financial Leasing, Trust \& Asset Management \\

\bottomrule
\end{tabularx}
\end{table}

%% file: tables/affiliation_distribution.tex
\begin{table*}[t]
\centering
\small
\setlength{\tabcolsep}{6pt}
\renewcommand{\arraystretch}{1.2}

\caption{Selected Institutional Affiliations of Experts Contributing to Subjective Tasks}
\label{tab:affiliation_distribution}

\begin{tabular}{p{3cm} p{8cm}}
\hline
\textbf{Sector} & \textbf{Institutions} \\
\hline

\textbf{Securities} & 
Huatai Securities, CITIC Securities, J Trust Global Securities\\

\textbf{Banking} & 
ICBC; China Development Bank; Nanyang Commercial Bank \\

\textbf{Asset Management} & 
Man Group, Value Partners Group, E Fund Management \\

\textbf{Investment Bank} & 
China International Capital Corporation\\

\textbf{Venture Capital} & 
Hongnuo Venture Capital\\

\textbf{Futures} & 
CITIC Futures \\

\textbf{Legal Services} & 
Dentons Shanghai Office \\

\hline
\end{tabular}
\end{table*}

%% file: tables/authoring_principles.tex
\begin{table}[htbp]
\centering
\small

\caption{Design Principles for Financial Deep Research Tasks}
\label{tab:authoring_principles}

\begin{tabular}{p{3cm} p{10cm}}
\hline
\textbf{Dimension} & \textbf{Description} \\
\hline

\textbf{Accuracy} & 
Problem statements must be clear, precise, and unambiguous, with explicitly defined constraints (e.g., time range, metrics, format). Outputs should be concise and standardized for consistent evaluation. \\

\textbf{Compliance} & 
Questions involving regulations or policies must rely on up-to-date and valid legal frameworks, ensuring correctness and regulatory compliance. \\

\textbf{Domain Relevance} & 
Questions should reflect real-world financial scenarios, using professional terminology and aligning with practical workflows such as research, risk analysis, and client management. \\

\textbf{Depth \& Complexity} & 
Tasks should require multi-step reasoning or cross-source analysis, going beyond simple lookup and reflecting realistic research difficulty. \\

\textbf{Scope \& Diversity} & 
Questions should cover diverse task types and global contexts, including variations in markets, regulations, standards, currencies, and analytical perspectives. \\

\hline
\end{tabular}
\end{table}

%% file: tables/task_schema.tex
\begin{table}[t]
\centering
\small
\setlength{\tabcolsep}{4pt}
\renewcommand{\arraystretch}{1.1}

\caption{Task Schema for Objective Questions}
\label{tab:task_schema}

\begin{tabularx}{\linewidth}{lX}
\toprule

\textbf{Field} & \textbf{Description} \\
\midrule

\multicolumn{2}{l}{\textbf{Identification}} \\
\emph{Task ID} & Unique identifier of the task. \\
\emph{Problem Statement} & The question or task description. \\
\emph{Language} & Language of the task, e.g., Chinese or English. \\
\emph{Classification} & Domain classification code and name . \\
\emph{Tags} & Array of keywords for task categorization (min 1). \\
\midrule

\multicolumn{2}{l}{\textbf{Answer Specification}} \\
\emph{Answer Type} & number, multi\_choice, short\_text. \\
\emph{Options} & For multiple-choice questions, the list of choices (min 5 when present). \\
\emph{Ground Truth} & The correct answer . \\
\emph{Format Prompt} & Instructions for answer formatting, if applicable. \\
\midrule

\multicolumn{2}{l}{\textbf{Difficulty \& Reasoning}} \\
\emph{Difficulty Level} & Level 1 (Easy), 2 (Medium), or 3 (Hard). \\
\emph{Number of Steps} & Number of reasoning steps required . \\
\emph{Step Details} & List of step-by-step reasoning descriptions (min 1). \\
\midrule

\multicolumn{2}{l}{\textbf{Tools \& Sources}} \\
\emph{Tools Required} & Search API, Web Browser, Multi-modality, Coding, File, or \texttt{N/A}. \\
\emph{Number of Tools} & Count of tools used. \\
\emph{Information Sources} & List of reference URLs or materials. \\
\emph{Source Count} & Number of referenced sources. \\
\midrule

\multicolumn{2}{l}{\textbf{Authorship \& Review}} \\
\emph{Author Name} & Name of the task designer. \\
\emph{Author Affiliation} & Institution of the author. \\
\emph{Status} & DRAFT, SUBMITTED, IN\_REVIEW, APPROVED, NEEDS\_REVISION, REJECTED, MERGED, or LOCKED. \\
\emph{Review Rounds} & Records of LLM and human expert reviews. \\
\bottomrule
\end{tabularx}
\end{table}

%% file: tables/rating_scale.tex
\begin{table}[t]
\centering
\small
\setlength{\tabcolsep}{4pt}
\renewcommand{\arraystretch}{1.1}

\caption{Recommendation Scoring Scheme for Task Quality Assessment}
\label{tab:rating_scale}

\begin{tabularx}{\linewidth}{l p{5cm} X}
\toprule

\textbf{Score} & \textbf{Label} & \textbf{Description} \\
\midrule

0 & Discard & The task is out of scope, lacks originality, is of low quality, or violates authoring principles. \\
1 & Uncertain, major revision needed & The task requires substantial modification, or the reviewer is uncertain about its quality. Please provide comments. \\
2 & Pending, minor revision needed & The task requires minor modifications. Please provide comments. \\
3 & Overly simplistic or artificially difficult & The task is too basic (easily answered by simple online search) or artificially difficult due to tool restrictions (e.g., heavy computation, rendering) that the evaluated models cannot use. \\
4 & Acceptable for candidate pool & The task is worth including but has minor flaws, such as high similarity to existing tasks, lack of business relevance, or has been solved by one or more models. \\
5 & High-quality for benchmark & The task exhibits complexity, realistic business scenarios, accurate answers, and correct formatting. It is suitable for the formal benchmark. \\
6 & Top-tier & Exceptional task, comparable to graduate or research-level quality. It deserves inclusion in the formal benchmark and can serve as a high-quality example. \\
\bottomrule
\end{tabularx}
\end{table}

%% file: tables/model_versions.tex
\begin{table}[t]
\centering
\small
\setlength{\tabcolsep}{6pt}
\renewcommand{\arraystretch}{1.1}

\caption{\textbf{Model and framework configurations used in experiments.} 
Release dates correspond to the publicly available model or system versions identified during evaluation. All experiments were conducted in April 2026 using the latest accessible versions at that time. For continuously updated proprietary Deep Research systems, actual deployed versions may differ from the publicly documented releases if silent updates were applied by providers.}
\label{tab:model_versions}

\begin{tabular}{lll} 
\toprule
\textbf{System} & \textbf{Model / System Version} & \textbf{Release Date} \\ 
\midrule

Gemini-deep-research         & Gemini-3-pro-preview & 2025.12.11 \\
o3-deep-research             & o3                   & 2025.2.2 \\
Perplexity-deep-research     & Llama 3.3 70B        & 2025.02.14 \\
Grok-3-deepsearch            & Grok 3               & 2025.02.19 \\
Doubao-deep-research         & --                   & 2025.06.30 \\
Qwen-deep-research           & --                   & 2025.12.15 \\
Kimi-deep-research           & Kimi-Researcher(trained on Kimi k1.5)    & 2025.06.20 \\
Jina-deepsearch              & --                   & -- \\
Tongyi-deepresearch-30b-a3b  & Tongyi-DeepResearch-30B-A3B   & 2025.11.05 \\
MiroThinker                  & MiroThinker-1.7      & 2026.03.11 \\
DeerFlow                     & --                   & 2026.04.15 \\
OpenClaw                     & v2026.4.8            & 2026.04.08 \\

\bottomrule
\end{tabular}
\end{table}

%% file: tables/agent_skills.tex
\begin{table}[h]
\centering
\small
\setlength{\tabcolsep}{6pt}
\renewcommand{\arraystretch}{1.1}

\caption{Agent skills employed in both the DeerFlow and OpenClaw frameworks}
\label{tab:agent_skills}

\begin{tabular}{l p{8.5cm}}
\toprule
\textbf{Skill} & \textbf{Description} \\
\midrule
Token \& Asset Management       & bankr, bankr-token-scam-analysis, stakr, hydrex, zyfai \\
Trading \& Market Intelligence  & signals, agenticbets, checkr, quotient, qrcoin \\
Financial Intelligence          & alphaear-news, alphaear-stock, alphaear-sentiment, alphaear-predictor, alphaear-signal-tracker, alphaear-logic-visualizer, alphaear-reporter, alphaear-search \\
Cross-chain \& DeFi             & trails, symbiosis, veil \\
Identity \& Reputation          & erc-8004, siwa, helixa, trustlayer-sybil-scanner, ens-primary-name \\
Social \& Messaging             & bankr-twitter-agent, botchan, neynar, productclank, yoink \\
Data \& Infrastructure          & quicknode, alchemy, zerion, darksol-random-oracle, onchainkit \\
Coordination \& Commerce        & nookplot, 0xwork, gitlawb, moltycash, endaoment, bankr-shopify \\
Mining \& Gaming                & BOTCOIN, litcoin, cattown \\
Security \& Privacy             & blueagent \\

\bottomrule
\end{tabular}
\end{table}

%% file: figures/score_comparison_heatmap.tex
\begin{figure}[ht]
  \centering
  \includegraphics[width=0.85\linewidth]{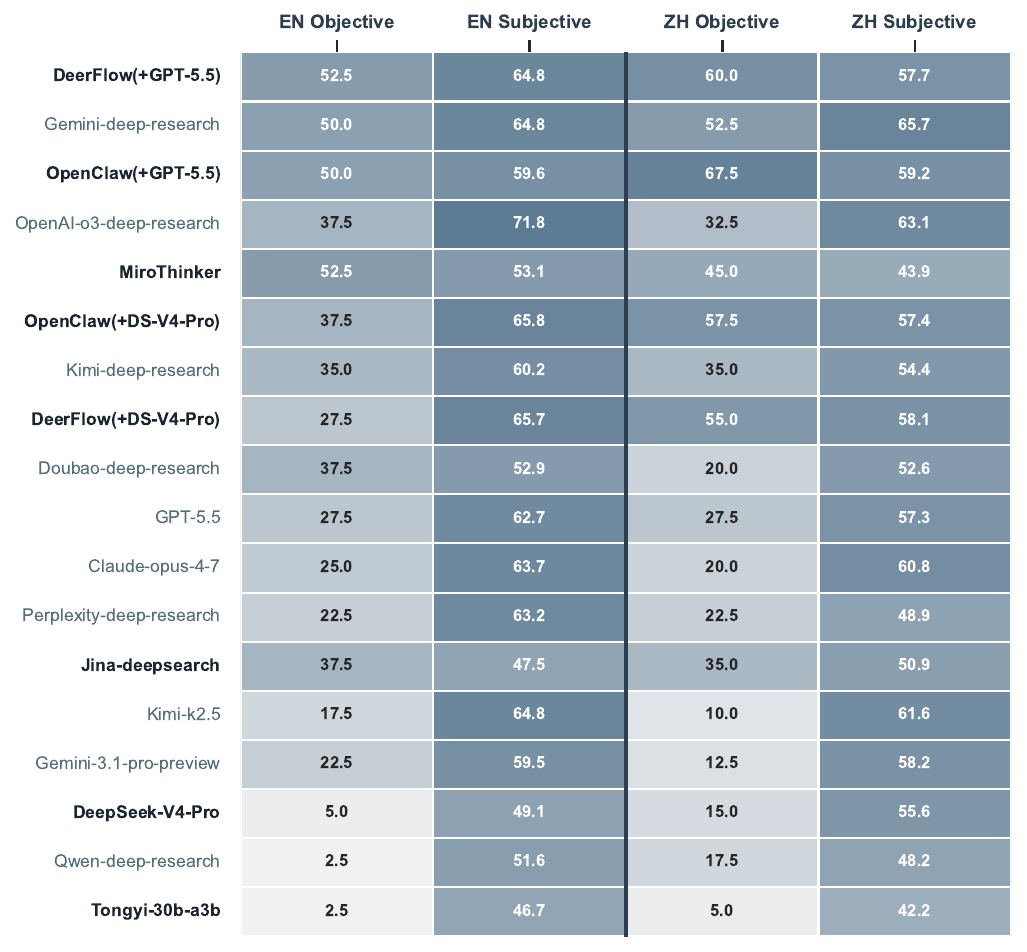} 
  \caption{\textbf{Skill-level diagnostic heatmap across ICBCBench dimensions.} This matrix illustrates the granular performance distribution across Objective reasoning and Subjective generation tasks. The vertical center line distinguishes between Global (EN) and Chinese (ZH) scenarios. The stark color contrast reveals a systemic difficulty gap in precise financial data extraction (Objective) compared to narrative report synthesis (Subjective). Model names in \textbf{bold} denote open-source systems.}
  \vspace{-0.3cm}
  \label{fig:score_comparison_heatmap}
\end{figure}

%% file: tables/results_subset_en.tex
\begin{table*}[t]
\centering
\small
\setlength{\tabcolsep}{5pt}
\renewcommand{\arraystretch}{1.1}

\caption{\textbf{Performance on English (EN) tasks in ICBCBench.} We report objective and subjective results for global market scenarios. Objective includes text-only and aggregated scores (All), while Subjective evaluates report quality via Expert Rubrics, Citation Consistency, and Source Quality. The Overall score is the arithmetic mean of the Objective and Subjective scores. The best and second-best scores are highlighted in \textbf{bold} and \underline{underline}, respectively. Higher is better for all metrics.}
\label{tab:results_subset_en}

\begin{tabular}{lcccccc}
\toprule
\multirow{2}{*}{\textbf{System}}
& \multicolumn{2}{c}{\textbf{Objective}}
& \multicolumn{3}{c}{\textbf{Subjective}}
& \multirow{2}{*}{\textbf{Overall}} \\
\cmidrule(lr){2-3} \cmidrule(lr){4-6}
& \textbf{Text-Only} & \textbf{All} & \textbf{Expert} & \textbf{Citation} & \textbf{Source} & \\
\midrule
\rowcolor{gray!30}
\multicolumn{7}{c}{\textbf{\textit{Closed}}} \\
Gemini-deep-research          & \underline{57.14} &           50.00 &           72.23 & \underline{56.94} &           12.86 & \underline{57.38} \\
OpenAI-o3-deep-research       &           37.14 &           37.50 &           78.55 &  \textbf{74.47} & \underline{15.49} &           54.67 \\
Kimi-deep-research            &           40.00 &           35.00 &           75.23 &            -- &            -- &           47.59 \\
Doubao-deep-research          &           42.86 &           37.50 &           66.17 &            -- &            -- &           45.22 \\
GPT-5.5                       &           20.00 &           27.50 &           78.37 &            -- &            -- &           45.09 \\
Claude-opus-4-7               &           17.14 &           25.00 &           79.63 &            -- &            -- &           44.36 \\
Perplexity-deep-research      &           25.71 &           22.50 &           78.97 &            -- &            -- &           42.84 \\
Gemini-3.1-pro-preview        &           20.00 &           22.50 &           74.42 &            -- &            -- &           41.02 \\
Grok-3-deepsearch             &            8.57 &           10.00 &           70.53 &            -- &            -- &           33.22 \\
Qwen-deep-research            &            2.86 &            2.50 &           64.48 &            -- &            -- &           27.05 \\
\midrule
\rowcolor{gray!30}
\multicolumn{7}{c}{\textbf{\textit{Open}}} \\
DeerFlow(+GPT-5.5)            & \underline{57.14} &  \textbf{52.50} &           81.07 &            -- &            -- &  \textbf{58.67} \\
OpenClaw(+GPT-5.5)            &           54.29 &           50.00 &           74.50 &            -- &            -- &           54.80 \\
MiroThinker                   &  \textbf{60.00} &  \textbf{52.50} &           66.43 &            -- &            -- &           52.83 \\
OpenClaw(+DeepSeek-V4-Pro)    &           40.00 &           37.50 &  \textbf{82.23} &            -- &            -- &           51.65 \\
DeerFlow(+DeepSeek-V4-Pro)    &           31.43 &           27.50 & \underline{82.13} &            -- &            -- &           46.60 \\
Jina-deepsearch               &           34.29 &           37.50 &           50.15 &           46.36 &  \textbf{27.54} &           42.50 \\
Kimi-k2.5                     &           14.29 &           17.50 &           81.02 &            -- &            -- &           41.16 \\
DeepSeek-V4-Pro               &            5.71 &            5.00 &           61.37 &            -- &            -- &           27.05 \\
Tongyi-deepresearch-30b-a3b   &            2.86 &            2.50 &           58.37 &            -- &            -- &           24.59 \\
\bottomrule
\end{tabular}
\end{table*}

%% file: tables/results_subset_zh.tex
\begin{table*}[t]
\centering
\small
\setlength{\tabcolsep}{5pt}
\renewcommand{\arraystretch}{1.1}

\caption{\textbf{Performance on Chinese (ZH) tasks in ICBCBench.} We report objective and subjective results for domestic market scenarios. Objective includes text-only and aggregated scores (All), while Subjective evaluates report quality via Expert Rubrics, Citation Consistency, and Source Quality. The Overall score is the arithmetic mean of the Objective and Subjective scores. The best and second-best scores are highlighted in \textbf{bold} and \underline{underline}, respectively. Higher is better for all metrics.}
\label{tab:results_subset_zh}

\begin{tabular}{lcccccc}
\toprule
\multirow{2}{*}{\textbf{System}}
& \multicolumn{2}{c}{\textbf{Objective}}
& \multicolumn{3}{c}{\textbf{Subjective}}
& \multirow{2}{*}{\textbf{Overall}} \\
\cmidrule(lr){2-3} \cmidrule(lr){4-6}
& \textbf{Text-Only} & \textbf{All} & \textbf{Expert} & \textbf{Citation} & \textbf{Source} & \\
\midrule
\rowcolor{gray!30}
\multicolumn{7}{c}{\textbf{\textit{Closed}}} \\
Gemini-deep-research          &           61.76 &           52.50 &           71.35 & \underline{73.16} &           12.97 & \underline{59.09} \\
OpenAI-o3-deep-research       &           38.24 &           32.50 &           66.65 &  \textbf{79.35} & \underline{18.65} &           47.81 \\
Kimi-deep-research            &           41.18 &           35.00 &           68.05 &            -- &            -- &           44.72 \\
GPT-5.5                       &           29.41 &           27.50 &           71.67 &            -- &            -- &           42.41 \\
Claude-opus-4-7               &           23.53 &           20.00 & \underline{76.03} &            -- &            -- &           40.41 \\
Doubao-deep-research          &           23.53 &           20.00 &           65.77 &            -- &            -- &           36.30 \\
Perplexity-deep-research      &           26.47 &           22.50 &           61.07 &            -- &            -- &           35.67 \\
Gemini-3.1-pro-preview        &           14.71 &           12.50 &           72.82 &            -- &            -- &           35.38 \\
Qwen-deep-research            &           20.59 &           17.50 &           60.32 &            -- &            -- &           32.88 \\
Grok-3-deepsearch             &            5.88 &            5.00 &           63.00 &            -- &            -- &           27.70 \\
\midrule
\rowcolor{gray!30}
\multicolumn{7}{c}{\textbf{\textit{Open}}} \\
OpenClaw(+GPT-5.5)            &  \textbf{70.59} &  \textbf{67.50} &           74.07 &            -- &            -- &  \textbf{63.38} \\
DeerFlow(+GPT-5.5)            & \underline{64.71} & \underline{60.00} &           72.08 &            -- &            -- &           58.84 \\
OpenClaw(+DeepSeek-V4-Pro)    &           61.76 &           57.50 &           71.70 &            -- &            -- &           57.43 \\
DeerFlow(+DeepSeek-V4-Pro)    &           61.76 &           55.00 &           72.60 &            -- &            -- &           56.54 \\
MiroThinker                   &           52.94 &           45.00 &           54.85 &            -- &            -- &           44.44 \\
Jina-deepsearch               &           41.18 &           35.00 &           54.88 &           38.02 &  \textbf{31.83} &           42.95 \\
Kimi-k2.5                     &           11.76 &           10.00 &  \textbf{77.00} &            -- &            -- &           35.80 \\
DeepSeek-V4-Pro               &           17.65 &           15.00 &           69.48 &            -- &            -- &           35.30 \\
Tongyi-deepresearch-30b-a3b   &            5.88 &            5.00 &           52.70 &            -- &            -- &           23.58 \\
\bottomrule
\end{tabular}
\end{table*}

%% file: tables/results_private_set.tex
\begin{table*}[t]
\centering
\small
\setlength{\tabcolsep}{5pt}
\renewcommand{\arraystretch}{1.08}

\caption{\textbf{Generalization performance on the private hold-out set of ICBCBench.} The private set is not publicly released and is designed to evaluate model generalization and prevent benchmark overfitting. Results are reported on both global (EN) and Chinese (ZH) scenarios across objective and subjective tasks. The best and second-best scores are highlighted in \textbf{bold} and \underline{underline}, respectively. Higher is better for all metrics.}
\label{tab:results_private_set}

\begin{tabular}{lcccccc}
\toprule
\multirow{2}{*}{\textbf{System}} & \multicolumn{3}{c}{\textbf{Global (EN)}} & \multicolumn{3}{c}{\textbf{Chinese (ZH)}} \\
\cmidrule(lr){2-4} \cmidrule(lr){5-7}
& \textbf{Objective} & \textbf{Subjective} & \textbf{Overall} & \textbf{Objective} & \textbf{Subjective} & \textbf{Overall} \\
\midrule
\rowcolor{gray!20}
\multicolumn{7}{c}{\textbf{\textit{Closed}}} \\
Gemini-deep-research                     & \underline{75.00} &           64.68 &           69.84 &           45.00 & \underline{63.19} &           54.09 \\
OpenAI-o3-deep-research                  &           55.00 &  \textbf{69.05} &           62.02 &           35.00 &           61.86 &           48.43 \\
Kimi-deep-research                       &           55.00 &           59.57 &           57.28 &           40.00 &           55.24 &           47.62 \\
Doubao-deep-research                     &           40.00 &           47.17 &           43.59 &           25.00 &           49.47 &           37.23 \\
Perplexity-deep-research                 &           20.00 &           60.53 &           40.27 &           30.00 &           45.64 &           37.82 \\
GPT-5.5                                  &           20.00 &           57.00 &           38.50 &           20.00 &           53.93 &           36.97 \\
Claude-opus-4-7                          &            5.00 &           62.57 &           33.78 &           20.00 &           58.38 &           39.19 \\
Gemini-3.1-pro-preview                   &           10.00 &           56.43 &           33.22 &           25.00 &           55.40 &           40.20 \\
Grok-3-deepsearch                        &           10.00 &           53.03 &           31.52 &           15.00 &           41.89 &           28.45 \\
Qwen-deep-research                       &           10.00 &           48.30 &           29.15 &           20.00 &           45.18 &           32.59 \\
\midrule
\rowcolor{gray!20}
\multicolumn{7}{c}{\textbf{\textit{Open}}} \\
OpenClaw(+DeepSeek-V4-Pro)               &  \textbf{85.00} &           64.83 &  \textbf{74.91} & \underline{55.00} &           56.09 & \underline{55.55} \\
DeerFlow(+DeepSeek-V4-Pro)               & \underline{75.00} & \underline{68.63} & \underline{71.81} &           40.00 &           56.27 &           48.14 \\
OpenClaw(+GPT-5.5)                       &           70.00 &           61.63 &           65.81 &  \textbf{60.00} &           57.24 &  \textbf{58.62} \\
DeerFlow(+GPT-5.5)                       &           65.00 &           59.80 &           62.40 &           45.00 &           57.67 &           51.34 \\
MiroThinker                              &           65.00 &           43.30 &           54.15 &           40.00 &           36.18 &           38.09 \\
Jina-deepsearch                          &           20.00 &           48.60 &           34.30 &           10.00 &           45.37 &           27.68 \\
Kimi-k2.5                                &            5.00 &           62.63 &           33.81 &           20.00 &  \textbf{64.16} &           42.08 \\
Tongyi-deepresearch-30b-a3b              &            0.00 &           46.27 &           23.14 &            5.00 &           36.91 &           20.95 \\
DeepSeek-V4-Pro                          &            5.00 &           22.50 &           13.75 &           20.00 &           49.47 &           34.73 \\
\bottomrule
\end{tabular}
\end{table*}

%% file: figures/correlation_accuracy_vs_rmsce_public.tex
\begin{figure}[t]
\centering
\includegraphics[width=0.85\linewidth]{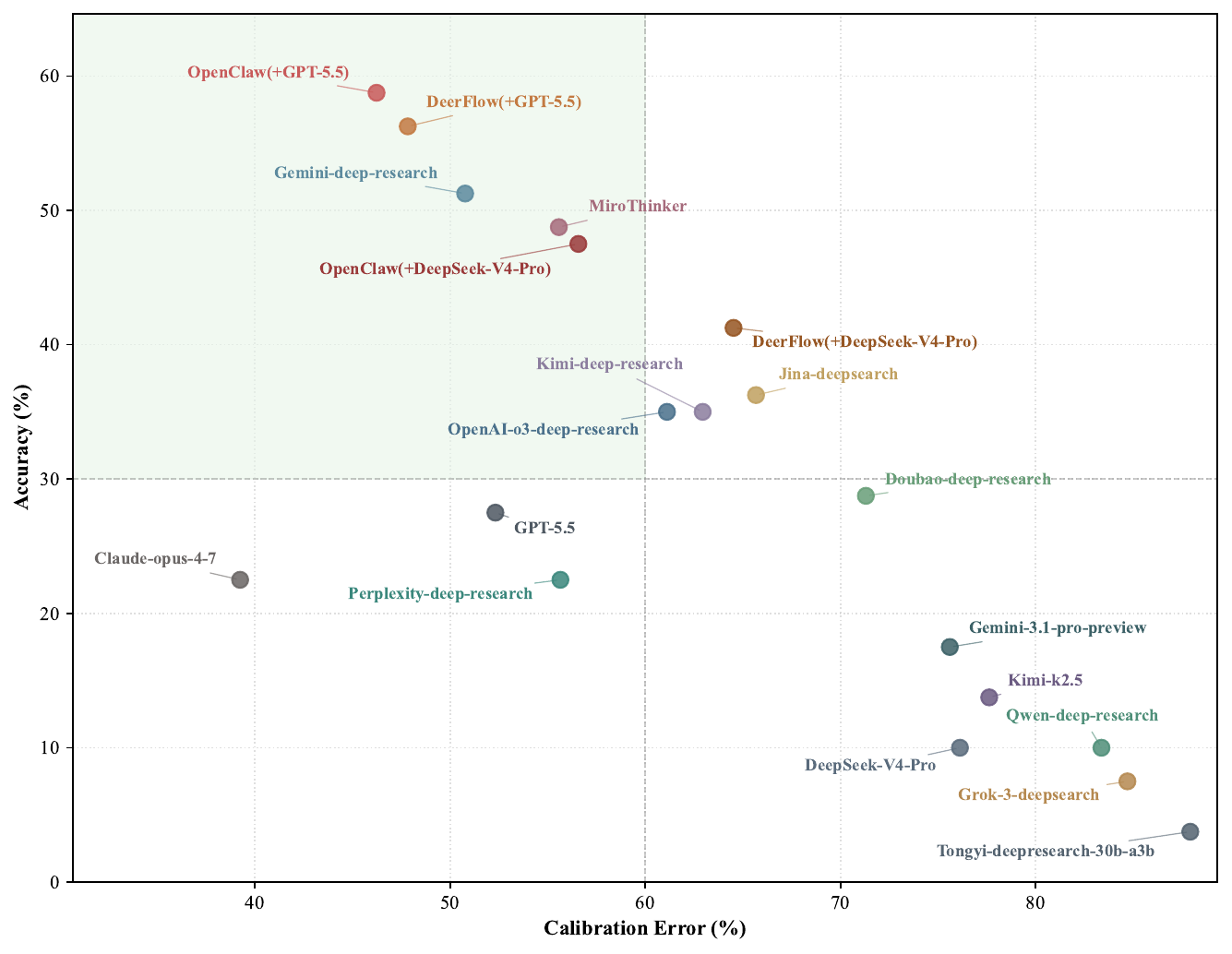}
\caption{\textbf{Accuracy versus calibration error on the objective subset.} Each point represents a model, with accuracy on the vertical axis and calibration error on the horizontal axis. The dashed reference lines mark accuracy = 30\% and calibration error = 60\%, respectively. The shaded green region in the top-left corner (accuracy $>$ 30\% and calibration error $<$ 60\%) highlights the ideal zone where models are both accurate and well-calibrated.}
\label{fig:correlation_accuracy_vs_rmsce_public}
\end{figure}

%% file: tables/objective_public_CalibErr.tex
\begin{table}[t]
\centering
\small
\caption{Objective Evaluation Results on Public Subset (All Languages). Higher Accuracy and lower Calibration Error are better.}
\label{tab:objective_public_CalibErr}
\begin{tabular}{lcc}
\toprule
\textbf{Model} & \textbf{Accuracy (\%)} & \textbf{Calibration Error (\%)} \\
\midrule
OpenClaw(+GPT-5.5) & 58.75 & 46.23 \\
DeerFlow(+GPT-5.5) & 56.25 & 47.83 \\
Gemini-deep-research & 51.25 & 50.77 \\
MiroThinker & 48.75 & 55.57 \\
OpenClaw(+DeepSeek-V4-Pro) & 47.50 & 56.57 \\
DeerFlow(+DeepSeek-V4-Pro) & 41.25 & 64.52 \\
Jina-deepsearch & 36.25 & 65.67 \\
OpenAI-o3-deep-research & 35.00 & 61.11 \\
Kimi-deep-research & 35.00 & 62.94 \\
Doubao-deep-research & 28.75 & 71.30 \\
GPT-5.5 & 27.50 & 52.32 \\
Claude-opus-4-7 & 22.50 & 39.24 \\
Perplexity-deep-research & 22.50 & 55.65 \\
Gemini-3.1-pro-preview & 17.50 & 75.60 \\
Kimi-k2.5 & 13.75 & 77.62 \\
DeepSeek-V4-Pro & 10.00 & 76.12 \\
Qwen-deep-research & 10.00 & 83.37 \\
Grok-3-deepsearch & 7.50 & 84.70 \\
Tongyi-deepresearch-30b-a3b & 3.75 & 87.92 \\
\bottomrule
\end{tabular}
\end{table}

%% file: tables/results_domain_specific.tex
\begin{table*}[t]
\centering
\small
\setlength{\tabcolsep}{6pt}
\renewcommand{\arraystretch}{1.1}

\caption{\textbf{Domain-specific objective performance across major financial sectors.} This table reports objective task accuracy across Banking, Capital Markets, Insurance, and Other Financial Services. The results reveal that Open-Agentic frameworks consistently outperform closed-source proprietary models across all domains, highlighting their robust factual extraction and tool-use capabilities in specialized financial contexts. The best and second-best scores are highlighted in \textbf{bold} and \underline{underline}.}

\label{tab:results_domain_specific}

\begin{tabular}{lccccc}
\toprule
\textbf{System} & \textbf{Banking} & \textbf{Capital Markets} & \textbf{Insurance} & \textbf{Others} & \textbf{Average} \\
\midrule

\rowcolor{gray!20}
\multicolumn{6}{c}{\textbf{\textit{Closed}}} \\
Gemini-deep-research          & \underline{53.57} & \underline{66.67} & 40.00 & 25.00 & 46.31 \\
Kimi-deep-research            & 28.57 & 50.00 & 30.00 & 25.00 & 33.39 \\
OpenAI-o3-deep-research       & 42.86 & 33.33 & 30.00 & 25.00 & 32.80 \\
GPT-5.5                       & 25.00 & 20.83 & 35.00 & 37.50 & 29.58 \\
Doubao-deep-research          & 35.71 & 29.17 & 25.00 & 12.50 & 25.60 \\
Perplexity-deep-research      & 17.86 & 25.00 & 25.00 & 25.00 & 23.21 \\
Claude-opus-4-7               & 28.57 & 12.50 & 30.00 & 12.50 & 20.89 \\
Gemini-3.1-pro-preview        & 21.43 & 16.67 & 15.00 & 12.50 & 16.40 \\
Grok-3-deepsearch             & 7.14 & 8.33 & 5.00 & 12.50 & 8.24 \\
Qwen-deep-research            & 10.71 & 20.83 & 0.00 & 0.00 & 7.89 \\
\rowcolor{gray!20}
\multicolumn{6}{c}{\textbf{\textit{Open}}} \\
OpenClaw(+GPT-5.5)            & \textbf{57.14} & 62.50 & \underline{50.00} & \textbf{75.00} & \textbf{61.16} \\
DeerFlow(+GPT-5.5)            & 46.43 & \underline{66.67} & \textbf{55.00} & \underline{62.50} & \underline{57.65} \\
MiroThinker                   & \underline{53.57} & 58.33 & 35.00 & 37.50 & 46.10 \\
OpenClaw(+DeepSeek-V4-Pro)    & 42.86 & \underline{66.67} & 35.00 & 37.50 & 45.51 \\
DeerFlow(+DeepSeek-V4-Pro)    & 21.43 & \textbf{70.83} & 25.00 & \underline{62.50} & 44.94 \\
Jina-deepsearch               & 39.29 & 41.67 & 30.00 & 25.00 & 33.99 \\
Kimi-k2.5                     & 14.29 & 12.50 & 20.00 & 0.00 & 11.70 \\
DeepSeek-V4-Pro               & 10.71 & 4.17 & 15.00 & 12.50 & 10.60 \\
Tongyi-deepresearch-30b-a3b   & 3.57 & 4.17 & 5.00 & 0.00 & 3.18 \\
\bottomrule
\end{tabular}
\end{table*}

%% file: tables/benchmark_comparison.tex
\begin{table}[t]
\centering
\small
\setlength{\tabcolsep}{3.5pt}
\renewcommand{\arraystretch}{1.05}

\caption{Comparison of ICBCBench with representative Deep Research benchmarks. 
\pmark\ denotes benchmarks where finance is a key domain. Answer-Based indicates tasks with well-defined, verifiable ground-truth answers. Citation Consistency refers to citation consistency verification, Source Authority evaluates the credibility of information sources, and Expert Rubrics denotes evaluation criteria curated with domain expert involvement.}
\label{tab:benchmark_comparison}

\begin{tabular}{>{\raggedright\arraybackslash}p{3.7cm}ccccccc}
\toprule

\textbf{Benchmark} &
\multicolumn{3}{c}{\textbf{Task Domain}} &
\multicolumn{4}{c}{\textbf{Evaluation}} \\

\cmidrule(lr){2-4} \cmidrule(lr){5-8}

& \textbf{Financial} & \textbf{Closed} & \textbf{Open}
& \makecell{\textbf{Answer}\\\textbf{Based}}
& \makecell{\textbf{Citation}\\\textbf{Consistency}}
& \makecell{\textbf{Source}\\\textbf{Authority}}
& \makecell{\textbf{Expert}\\\textbf{Rubrics}} \\

\midrule

HLE~\cite{phan2025humanity}          & \xmark & \cmark & \xmark & \cmark & \xmark & \xmark & \xmark \\
GAIA~\cite{mialon2023gaia}          & \xmark & \cmark & \xmark & \cmark & \xmark & \xmark & \xmark \\
BrowseComp~\cite{Wei2025BrowseCompAS}          & \xmark & \cmark & \xmark & \cmark & \xmark & \xmark & \xmark \\
DeepSearchQA~\cite{Gupta2026DeepSearchQABT}          & \pmark & \cmark & \xmark & \cmark & \xmark & \xmark & \xmark \\
FinSearchComp~\cite{Hu2025FinSearchCompTA}    & \cmark & \cmark & \xmark & \cmark & \xmark & \xmark & \xmark \\
FinGAIA~\cite{Zeng2025FinGAIAAC}    & \cmark & \cmark & \xmark & \cmark & \xmark & \xmark & \xmark \\
FinDeepForecast~\cite{Li2026FinDeepForecastAL}    & \cmark & \cmark & \xmark & \cmark & \xmark & \xmark & \xmark \\
DeepResearch Bench~\cite{Du2025DeepResearchBA}  & \pmark & \xmark & \cmark & \xmark & \cmark & \xmark & \xmark \\
DR. BENCH~\cite{Yao2025DrBA}          & \pmark & \xmark & \cmark & \xmark & \cmark & \cmark & \cmark \\
LiveResearchBench~\cite{wang2025liveresearchbenchlivebenchmarkusercentric}   & \pmark & \xmark & \cmark & \xmark & \cmark & \xmark & \cmark \\
ResearchRubrics~\cite{Sharma2025ResearchRubricsAB}          & \pmark & \xmark & \cmark & \xmark & \xmark & \xmark & \cmark \\
DEER~\cite{han2026deerbenchmarkevaluatingdeep}          & \pmark & \xmark & \cmark & \xmark & \cmark & \cmark & \cmark \\
DRBench~\cite{abaskohi2026drbenchrealisticbenchmarkenterprise}          & \xmark & \xmark & \cmark & \xmark & \cmark & \xmark & \xmark \\
DRACO~\cite{Zhong2026DRACOAC}          & \pmark & \xmark & \cmark & \xmark & \cmark & \xmark & \cmark \\
MiroEval~\cite{ye2026miroevalbenchmarkingmultimodaldeep}        & \pmark & \xmark & \cmark & \xmark & \cmark & \xmark & \xmark \\
FinRpt~\cite{Jin2025FinRptDE}    & \cmark & \xmark & \cmark & \xmark & \xmark & \xmark & \xmark \\
FinResearchBench~\cite{Sun2025FinResearchBenchAL}    & \cmark & \xmark & \cmark & \xmark & \xmark & \xmark & \xmark \\

\midrule

\textbf{ICBCBench (Ours)} &
\textbf{\cmark} &
\textbf{\cmark} &
\textbf{\cmark} &
\textbf{\cmark} &
\textbf{\cmark} &
\textbf{\cmark} &
\textbf{\cmark} \\

\bottomrule
\end{tabular}
\vspace{-0.3cm}
\end{table}

%% file: prompts/appendix_case_objective.tex
\begin{figure}[ht]
\centering
\begin{tcolorbox}[
    enhanced,
    width=\linewidth,
    colback=black!3,        
    colframe=black!40,      
    colbacktitle=black!60,  
    coltitle=white,         
    fonttitle=\bfseries,    
    title={\footnotesize Case Studies: Objective Tasks},
    boxrule=0.6pt,
    arc=2mm,                
    left=2mm, right=2mm, top=2mm, bottom=2mm, 
    boxsep=1mm,
    titlerule=0pt,           
    bottomrule at break=0pt, 
    toprule at break=0pt,    
    pad at break=2mm         
]
\footnotesize 
\linespread{1.1}\selectfont 

{\color{gray}\footnotesize [Example 1: Short Answer (Banking)]} \par \vspace{0.4em}
\textbf{Question:}\\
You are a Senior M\&A Strategist in the Capital Markets Division of an institution. In its research report dated December 18, 2025, Barclays listed Allfunds, a fund distribution platform, as one of the core assets that BNP Paribas (BNPP) plans to divest to offset the capital consumption arising from the acquisition of Athlon. The divestment of Allfunds will make a significant marginal contribution to whether BNPP can ultimately achieve its 13\% CET1 target.

Based on the forward-looking forecasts in the Barclays research report and the third-party cross-border transaction legal records released on January 22, 2026:
\begin{enumerate}[label=\textbf{\arabic*.}, leftmargin=1.5em, itemsep=0.2em, parsep=0pt, topsep=0.4em]
    \item State the name of the European stock exchange operator that officially announced a Recommended Acquisition for Allfunds at an approximate valuation of €5.3 billion on January 21, 2026.
    \item According to the operational fundamentals disclosed in the acquisition agreement, what is the all-time high figure of Assets under Administration (AuA) achieved by Allfunds on the eve of the acquisition (i.e., as of September 30, 2025)? (Provide the numerical value in trillion euros, rounded to one decimal place.)
\end{enumerate}
\vspace{0.4em}
\textbf{Ground Truth:} \\
Deutsche Börse; 1.7

\vspace{1em}
{\color{gray}\footnotesize [Example 2: Multiple Choice (Insurance)]} \par \vspace{0.4em}
\textbf{Question:}\\
Allstate Corporation's performance in the first half of 2025 demonstrated a coexistence of strong profitability and significant catastrophe losses. Net income increased substantially in the second quarter, while catastrophe losses in April and May were particularly notable. Based on public reports regarding Allstate in 2025, which of the following statements are correct? (Multiple answers possible)
\begin{enumerate}[label=\textbf{\Alph*.}, leftmargin=1.5em, itemsep=0.4em, parsep=0pt, topsep=0.4em]
    \item In Q2 2025, Allstate's Property-Liability underlying combined ratio was 79.5\%, representing a significant improvement compared to the same period last year.
    \item In May 2025, Allstate raised the top of its catastrophe reinsurance tower to \$9.5 billion of coverage, with a retention of \$1 billion.
    \item The pre-tax total catastrophe losses for April and May 2025 combined were \$1.37 billion, of which approximately 70\% resulted from three widespread wind and hail events.
    \item Despite total catastrophe losses nearing \$2 billion in Q2 2025, Allstate's net income reached \$2.1 billion, significantly higher than the prior year period.
    \item In July 2025, Allstate estimated catastrophe losses of \$184 million, which entirely resulted from 19 wind and hail events and were lower than the level in July 2024.
    \item In Q1 2025, the company's net catastrophe losses decreased to \$2.2 billion due to \$1.1 billion in reinsurance recoveries.
    \item In Q3 2025, the underwriting profit for Allstate's homeowners insurance business climbed to \$1.1 billion, primarily benefiting from lower catastrophe losses.
\end{enumerate}
\vspace{0.4em}
\textbf{Ground Truth:} \\
A, D, F

\end{tcolorbox}
\caption{Illustrative examples of objective tasks in ICBCBench, spanning banking and insurance domains, requiring precise data extraction and multi-hop reasoning over financial reports.}
\label{fig:appendix_case_objective}
\end{figure}

%% file: prompts/appendix_case_subjective.tex
\begin{figure}[ht]
\centering
\begin{tcolorbox}[
    enhanced,
    width=\linewidth,
    colback=black!3,       
    colframe=black!40,     
    colbacktitle=black!60,  
    coltitle=white,         
    fonttitle=\bfseries,    
    title={\footnotesize Case Study: Subjective Tasks},
    boxrule=0.6pt,          
    arc=2mm,                
    left=2mm, right=2mm, top=2mm, bottom=2mm, 
    boxsep=1mm,
    titlerule=0pt,           
    bottomrule at break=0pt, 
    toprule at break=0pt,    
    pad at break=2mm         
]
\footnotesize 
\linespread{1.1}\selectfont 

{\color{gray}\footnotesize [Example 1: Banking Digital Operations Research]} \par \vspace{0.4em}
{\color{gray}\footnotesize [Context]} \\
With the continuous advancement of banking digitalization and financial technology, digital operations have become a core capability connecting customers, products, channels, and services. Centered on the customer lifecycle, banks are gradually building digital operation systems characterized by data-driven decision-making, process automation, and refined management. However, the actual effectiveness and inherent constraints of these systems remain to be systematically evaluated.

\vspace{0.8em}
{\color{gray}\footnotesize [Task Requirement]} \\
Please conduct an analysis of the banking digital operation system and, based on concrete case studies, produce a structured research report addressing the following aspects:

\begin{enumerate}[label=\textbf{\arabic*.}, leftmargin=1.5em, itemsep=0.4em, parsep=0pt]
    \item \textbf{Current Practices and Implementation:} 
    Systematically review current digital operation practices across key stages such as customer acquisition, activation, conversion, and retention. Analyze how data-driven approaches, process automation, and refined management are applied in these contexts.
    
    \item \textbf{Effectiveness and Challenges:} 
    Evaluate the effectiveness of these digital operation practices in improving customer value, operational efficiency, and service experience, and identify the main constraints and challenges they face.
    
    \item \textbf{Optimization Pathways and Future Directions:} 
    Considering technological advancements and evolving customer behavior, propose optimization directions and actionable pathways for future banking digital operation models from the perspectives of operational models, organizational mechanisms, and system capabilities.
\end{enumerate}

{\color{gray}\footnotesize [Example 2: AI Chip Industry Transformation Analysis]} \par \vspace{0.4em}
{\color{gray}\footnotesize [Context]} \\
Based on the current computational principles of large language model inference, the direction of model evolution, and the demand characteristics of downstream customers, please study how the large-scale deployment of model inference may drive the evolution of AI chip architectures, including but not limited to Processing-Near-Memory and 3D memory-logic stacking, changes in chip/server interconnect methods, and adjustments in model architectures. Further analyze which specific industry sectors may be affected by these changes.

\vspace{0.8em}
{\color{gray}\footnotesize [Task Requirement]} \\
Please follow the requirements below:

\begin{enumerate}[label=\textbf{\arabic*.}, leftmargin=1.5em, itemsep=0.4em, parsep=0pt]
    \item \textbf{Technical Pipeline and Bottlenecks:}
    From a technical perspective, conduct a comprehensive study of the full operational pipeline of current large-model inference services, including but not limited to model computation mechanisms, how hardware infrastructure performs the corresponding data processing, and storage/communication mechanisms. Under current trends in inference demand, identify the technical bottlenecks that may arise at each stage and discuss potential future improvement directions. All technical arguments should be supported by rigorously verified data sources to ensure the analysis is accurate, comprehensive, and up to date.
    
    \item \textbf{Industrial Impact and Competitive Landscape:}
    For each major technological improvement direction, assess the industrial beneficiaries and adversely affected parties, including both directly impacted stakeholders and important indirectly affected stakeholders. Analyze the channels of impact, transmission speed, and potential competitive dynamics across industries, and based on this, evaluate the likely industrial landscape over the next two to three years.
\end{enumerate}

\end{tcolorbox}
\caption{Illustrative examples of subjective research tasks in the ICBCBench, requiring multi-dimensional analysis and structured reporting.}
\label{fig:appendix_case_subjective}
\end{figure}

%% file: prompts/appendix_case_expert_rubric.tex
\begin{figure}[ht]
\centering
\begin{tcolorbox}[
    enhanced,
    width=\linewidth,
    colback=black!3,        
    colframe=black!40,      
    colbacktitle=black!60,  
    coltitle=white,         
    fonttitle=\bfseries,    
    title={\footnotesize Case Study: Subjective Task Expert Rubric (Banking Digital Operations Research)},
    boxrule=0.6pt,
    arc=2mm,                
    left=2mm, right=2mm, top=2mm, bottom=2mm, 
    boxsep=1mm,
    titlerule=0pt,           
    bottomrule at break=0pt, 
    toprule at break=0pt,    
    pad at break=2mm         
]
\footnotesize 
\linespread{1.1}\selectfont 

{\color{gray}\footnotesize [Rubric Overview]} \\
This framework provides a quantitative assessment (100 pts) across 4 primary and 12 secondary dimensions, emphasizing lifecycle closed-loops, implementation granularity, and risk awareness.

\vspace{0.6em}
{\color{gray}\footnotesize [Dimensions \& Detailed Criteria]}

\begin{enumerate}[label=\textbf{\arabic*.}, leftmargin=1.5em, itemsep=0.8em, parsep=0pt]
    \item \textbf{Framework \& Current-Practice Mapping (30 pts)}
    \begin{itemize}[leftmargin=1em, nosep, itemsep=0.2em]
        \item \textit{Closed-Loop Completeness (10 pts):} Mapping journeys (Data $\to$ Insight $\to$ Outreach $\to$ Conversion $\to$ Retention) with clear stage definitions.
        \item \textit{Acquisition Depth (10 pts):} Funnel decomposition (Spend $\to$ Reach $\to$ KYC $\to$ Active) and quality control (CAC, anti-fraud).
        \item \textit{Retention \& Loyalty (10 pts):} Trigger mechanisms for churn warning and product-specific strategies (Deposits, Wealth, Credit Cards).
    \end{itemize}

    \item \textbf{Data-Driven \& Fine-Grained Effectiveness (25 pts)}
    \begin{itemize}[leftmargin=1em, nosep, itemsep=0.2em]
        \item \textit{Data Foundation (8 pts):} Unified ID, data governance, and compliance boundaries (minimization, authorization audit trails).
        \item \textit{KPI Verifiability (9 pts):} Multi-layered metrics (Outcome/Process/Risk) and causal methods (A/B testing, cohort/attribution analysis).
        \item \textit{Reproducibility (8 pts):} Strategy granularity including frequency rules, cost constraints, and trade-offs between short-term vs. LTV.
    \end{itemize}

    \item \textbf{System Capabilities \& Organization (20 pts)}
    \begin{itemize}[leftmargin=1em, nosep, itemsep=0.2em]
        \item \textit{Automation Value (8 pts):} Impact on SLA and productivity via automated journey triggers and ticket/work-order routing.
        \item \textit{Architecture Match (7 pts):} CDP, CRM, and recommendation engine integration with legacy core systems and real-time constraints.
        \item \textit{Organizational Mechanisms (5 pts):} Approval workflows, HQ/Branch role division, and incentive balancing (Scale vs. Quality).
    \end{itemize}

    \item \textbf{Gap Diagnosis \& Optimization Path (25 pts)}
    \begin{itemize}[leftmargin=1em, nosep, itemsep=0.2em]
        \item \textit{Constraint Diagnosis (8 pts):} Root-cause analysis of data silos, model bias, outreach noise, and compliance restrictions.
        \item \textit{Tech-Trend Linking (8 pts):} Actionable landing points for AIGC, privacy computing, and event-driven operations.
        \item \textit{Implementation Roadmap (9 pts):} Phased objectives (0--3/3--6/6--12 months) with resource dependencies and "what-first" logic.
    \end{itemize}
\end{enumerate}

{\color{gray}\footnotesize [Scoring Example (10-point scale)]} \par \vspace{0.4em}
\textbf{9--10 pts (High):} Builds a clear, executable closed loop with precise input/output and metric definitions for each stage of the financial journey. \\
\textbf{6--8 pts (Medium):} Framework is largely complete but lacks detail in feedback iterations or specific post-loan/investment operation steps. \\
\textbf{0--5 pts (Low):} Fragmented framework limited to concept listing; lacks an end-to-end journey or an executable logic structure.

\end{tcolorbox}
\caption{Comprehensive scoring rubric for subjective financial research reports, reflecting all 12 secondary dimensions used in ICBCBench.}
\label{fig:appendix_case_expert_rubric}
\end{figure}

%% file: prompts/query_refinement_prompt.tex
\begin{figure}[t]
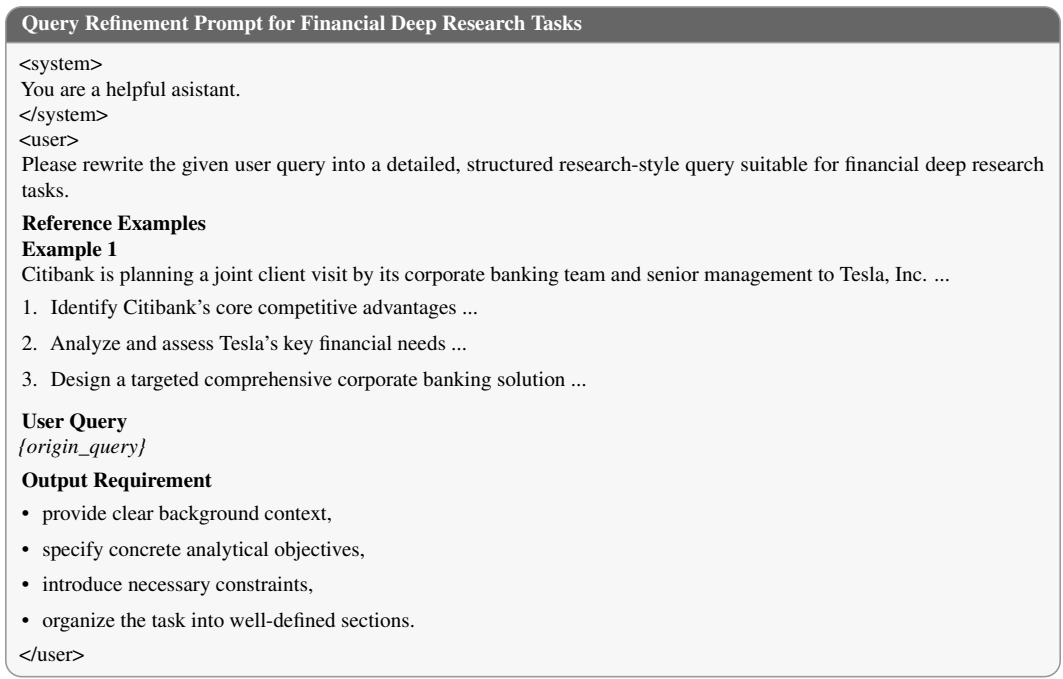

\centering

\begin{tcolorbox}[
    enhanced,
    width=\linewidth,
    colback=black!3,
    colframe=black!40,
    colbacktitle=black!60,
    coltitle=white,
    fonttitle=\bfseries,
    title={\footnotesize Query Refinement Prompt for Financial Deep Research Tasks},
    boxrule=0.6pt,
    arc=2mm,
    left=1mm,
    right=1mm,
    top=1mm,
    bottom=1mm,
    boxsep=1mm
]

\footnotesize

<system>\\
You are a helpful asistant.\\
</system>\\
<user>\\
Please rewrite the given user query into a detailed, structured research-style query suitable for financial deep research tasks.

\vspace{0.4em}
\textbf{Reference Examples}

\textbf{Example 1}

Citibank is planning a joint client visit by its corporate banking team and senior management to Tesla, Inc. ...

\begin{enumerate}[leftmargin=*, itemsep=2pt, topsep=2pt]
    \item Identify Citibank’s core competitive advantages ...
    \item Analyze and assess Tesla’s key financial needs ...
    \item Design a targeted comprehensive corporate banking solution ...
\end{enumerate}

\vspace{0.4em}
\textbf{User Query}

\textit{\{origin\_query\}}

\vspace{0.4em}
\textbf{Output Requirement}

\begin{itemize}[leftmargin=*, itemsep=2pt, topsep=2pt]
    \item provide clear background context,
    \item specify concrete analytical objectives,
    \item introduce necessary constraints,
    \item organize the task into well-defined sections.
\end{itemize}
</user>

\end{tcolorbox}

\caption{Query refinement prompt used to transform raw user queries into structured research tasks.}
\label{fig:query_refinement_prompt}

\end{figure}

%% file: prompts/objective_evaluation_prompt.tex
\begin{figure}[t]
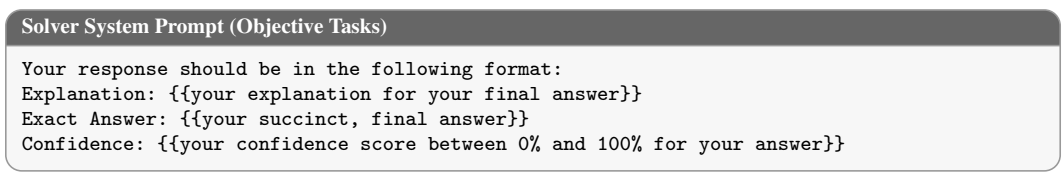

\centering
\begin{tcolorbox}[
    enhanced,
    width=\linewidth,
    colback=black!3,
    colframe=black!40,
    colbacktitle=black!60,
    coltitle=white,
    fonttitle=\bfseries,
    title={\footnotesize Solver System Prompt (Objective Tasks)},
    boxrule=0.6pt,
    arc=2mm,
    left=1mm,
    right=1mm,
    top=1mm,
    bottom=1mm,
    boxsep=1mm
]
\footnotesize
\begin{verbatim}
Your response should be in the following format:
Explanation: {{your explanation for your final answer}}
Exact Answer: {{your succinct, final answer}}
Confidence: {{your confidence score between 0% and 100% for your answer}}
\end{verbatim}
\end{tcolorbox}
\caption{Solver prompt used for objective tasks.}
\label{fig:solver_prompt}
\end{figure}

\begin{figure}[t]
\centering
\begin{tcolorbox}[
    enhanced,
    width=\linewidth,
    colback=black!3,
    colframe=black!40,
    colbacktitle=black!60,
    coltitle=white,
    fonttitle=\bfseries,
    title={\footnotesize Judge Prompt (Objective Tasks)},
    boxrule=0.6pt,
    arc=2mm,
    left=1mm,
    right=1mm,
    top=1mm,
    bottom=1mm,
    boxsep=1mm
]
\footnotesize
\begin{verbatim}
Judge whether the following [response] to [question] is correct or not based on the precise 
and unambiguous [correct_answer] below.

[question]: {question}

[response]: {response}

[correct_answer]: {correct_answer}

Your judgement must be output strictly in valid JSON format and must contain the following 
keys. Please adhere to the data types and dynamic criteria specified for each field:

{{
    "extracted_final_answer": (string) The final exact answer extracted from the [response]. 
    Put "None" if there is no exact, final answer to extract.
    
    "reasoning": (string) Explain why the extracted_final_answer is correct or incorrect 
    based on [correct_answer], focusing only on if there are meaningful differences between 
    [correct_answer] and the extracted_final_answer. Do not comment on any background to 
    the problem, do not attempt to solve the problem, do not argue for any answer different 
    than [correct_answer], focus only on whether the answers match.
    
    "correct": (string) Answer "yes" if extracted_final_answer matches the [correct_answer] 
    given above, or is within a small margin of error for numerical problems. Answer "no" 
    otherwise (i.e., if there is any inconsistency, ambiguity, non-equivalency, or if the 
    extracted answer is incorrect).
    
    "confidence": (integer) The extracted confidence score as a number between 0 and 100 
    from the [response]. Put 100 if there is no confidence score available.
}}
\end{verbatim}
\end{tcolorbox}
\caption{Judge prompt used to evaluate solver responses against ground truth.}
\label{fig:judge_prompt}
\end{figure}

%% file: prompts/subjective_evaluation_prompt.tex
\begin{figure}[t]
\centering
\begin{tcolorbox}[
    enhanced,
    width=\linewidth,
    colback=black!3,
    colframe=black!40,
    colbacktitle=black!60,
    coltitle=white,
    fonttitle=\bfseries,
    title={\footnotesize Judge Prompt (Subjective Tasks)},
    boxrule=0.6pt,
    arc=2mm,
    left=1mm,
    right=1mm,
    top=1mm,
    bottom=1mm,
    boxsep=1mm
]
\footnotesize
\begin{verbatim}
You are an extremely strict and fastidious senior review expert for financial research 
reports. Your task is to carefully read the [Research Report] and conduct an **extremely 
rigorous** quantitative evaluation against the [Grading Criteria].
### Assessment Principles
1. Evidence-Driven: Every point awarded must be supported by specific data, factual 
citations, or complete logical chains found in the report. Merely "mentioning a concept" 
earns no points.
2. High-Score Threshold: Top-tier scores require the report content in that section to 
possess an extremely high information density and align perfectly with the grading criteria. 
Content lacking depth must be restricted to the middle or lowest scoring tiers.
3. Granularity Check: If the report merely provides broad descriptions of phenomena without 
structured breakdowns and substantive argumentation, it should be deemed as lacking support. 
You must differentiate between "surface-level statements" and "in-depth analysis" based on 
the grading criteria.
### Evaluation Process
To ensure the objectivity and accuracy of your scoring, please execute the following 
evaluation steps internally, but ultimately output ONLY the JSON result:
1. Extract Text: For each secondary dimension, first locate the corresponding text content 
in the full report, such as specific facts, data, or logical chains. Ensure the extracted 
content is closely related to the dimension and can objectively support your score.
2. Identify Shortcomings: Objectively analyze whether the argumentation under that 
dimension forms a closed loop. Point out specific areas in the report where there is a 
lack of evidence, logical gaps, or insufficient granularity.
3. Align with Tier: Strictly compare the extracted text content with the tier descriptions 
(e.g., 7-8 points, 4-6 points) under that dimension to determine which tier the report 
falls into.
4. Precise Scoring: Within the determined tier range, assign a specific score based on the 
comprehensiveness of the evidence, and write a brief rationale for the points awarded or 
deducted.
### Grading Criteria
{criteria_text}
### Research Report
{report_text}

### Output Requirements
Please directly output a valid JSON array containing the analysis and score for each 
dimension. 
Do not include any Markdown formatting symbols (such as ```json or ```).
[
    {{
      "dimension_id": "1.1",
      "dimension_name": "Name of the secondary dimension",
      "evidence": "Objective supporting information found in the report",
      "shortcomings": "From a strict perspective, point out specific flaws where the report 
      is not fully articulated, missing information, or speaking in generalities",
      "reasoning": "Combining the shortcomings above, provide the rationale for point 
      deductions and tier placement",
      "score": 0
    }},
    {{
      "dimension_id": "1.2",
      "dimension_name": "Name of the secondary dimension",
      "evidence": "...",
      "shortcomings": "...",
      "reasoning": "...",
      "score": 0
    }}
    // ... Please ensure you output the evaluation for ALL secondary dimensions completely, 
    keeping the structure consistent ...
]
\end{verbatim}
\end{tcolorbox}
\caption{Judge prompt used to evaluate reports through criteria from experts.}
\label{fig:subjective_judge_prompt}
\end{figure}

%% file: references.bib
@misc{openai2024deepresearch,
  author = {{OpenAI}},
  title = {Introducing Deep Research},
  year = {2024},
  howpublished = {\url{https://openai.com/index/introducing-deep-research/}},
  note = {Accessed: 2026-03-13}
}

@misc{google2024geminideepresearch,
  author = {{Google}},
  title = {Try Deep Research and our new experimental model in Gemini, your AI assistant},
  year = {2024},
  howpublished = {\url{https://blog.google/products-and-platforms/products/gemini/google-gemini-deep-research/}},
  note = {Accessed: 2026-03-13}
}

@misc{perplexity2025deepresearch,
  author = {{Perplexity AI}},
  title = {Introducing Perplexity Deep Research},
  year = {2025},
  howpublished = {\url{https://www.perplexity.ai/hub/blog/introducing-perplexity-deep-research}},
  note = {Accessed: 2026-03-13}
}

@misc{openai2025o3deepresearch,
  author = {{OpenAI}},
  title = {o3-deep-research model},
  year = {2025},
  url = {https://platform.openai.com/docs/models/o3-deep-research},
  note = {OpenAI API documentation, accessed 2026-04-18}
}

@misc{xai2025grok3,
  author = {{xAI}},
  title = {Grok 3 Beta --- The Age of Reasoning Agents},
  year = {2025},
  url = {https://x.ai/news/grok-3},
  note = {Accessed: 2026-03-13}
}

@misc{doubao_chat_2026,
  author = {{ByteDance}},
  title = {Doubao Chat},
  year = {2026},
  url = {https://www.doubao.com/chat/},
  note = {Accessed: 2026-03-13}
}

@misc{qwen2025deepresearch,
  author = {{Qwen Team}},
  title = {Qwen DeepResearch: When Inspiration Becomes Its Own Execution},
  year = {2025},
  url = {https://qwen.ai/blog?id=qwen-deepresearch},
  note = {Accessed: 2026-03-13}
}

@misc{moonshot2025kimi_researcher,
  author = {{Moonshot AI}},
  title = {Kimi Researcher: End-to-End RL Training for Deep Research Agents},
  year = {2025},
  url = {https://moonshotai.github.io/Kimi-Researcher/},
  note = {Accessed: 2026-03-13}
}

@misc{google2025gemini3,
  author = {{Google}},
  title = {A New Era of Intelligence with Gemini 3},
  year = {2025},
  url = {https://blog.google/products-and-platforms/products/gemini/gemini-3/},
  note = {Accessed: 2026-03-13}
}

@misc{googledeepmind2025gemini31pro,
  title        = {Gemini 3.1 Pro Model Card},
  author       = {{Google DeepMind}},
  year         = {2025},
  howpublished = {\url{https://deepmind.google/models/model-cards/gemini-3-1-pro/}},
  note         = {Accessed: 2026-03-19}
}

@misc{openai2025gpt54,
  title        = {Introducing GPT-5.4},
  author       = {{OpenAI}},
  year         = {2025},
  howpublished = {\url{https://openai.com/index/introducing-gpt-5-4/}},
  note         = {Accessed: 2026-03-19}
}

@misc{openai2026gpt55,
  author = {{OpenAI}},
  title = {Introducing GPT-5.5},
  year = {2026},
  url = {https://openai.com/index/introducing-gpt-5-5/},
  note = {Accessed: 2026-04-24}
}

@misc{anthropic2026opus47,
  author = {{Anthropic}},
  title = {Introducing Claude Opus 4.7},
  year = {2026},
  url = {https://www.anthropic.com/news/claude-opus-4-7},
  note = {Accessed: 2026-03-13}
}

@article{team2026kimi,
  title={Kimi K2. 5: Visual Agentic Intelligence},
  author={Team, Kimi and Bai, Tongtong and Bai, Yifan and Bao, Yiping and Cai, SH and Cao, Yuan and Charles, Y and Che, HS and Chen, Cheng and Chen, Guanduo and others},
  journal={arXiv preprint arXiv:2602.02276},
  year={2026}
}

@misc{jina2025deepsearch,
  author = {{Jina AI}},
  title = {Jina DeepSearch},
  year = {2025},
  url = {https://jina.ai/deepsearch/},
  note = {Accessed: 2026-03-13}
}

@article{team2025tongyi,
  title={Tongyi deepresearch technical report},
  author={Team, Tongyi DeepResearch and Li, Baixuan and Zhang, Bo and Zhang, Dingchu and Huang, Fei and Li, Guangyu and Chen, Guoxin and Yin, Huifeng and Wu, Jialong and Zhou, Jingren and others},
  journal={arXiv preprint arXiv:2510.24701},
  year={2025}
}

@article{team2025mirothinker,
  title={Mirothinker: Pushing the performance boundaries of open-source research agents via model, context, and interactive scaling},
  author={Team, MiroMind and Bai, Song and Bing, Lidong and Chen, Carson and Chen, Guanzheng and Chen, Yuntao and Chen, Zhe and Chen, Ziyi and Dai, Jifeng and Dong, Xuan and others},
  journal={arXiv preprint arXiv:2511.11793},
  year={2025}
}

@misc{bytedance2026deerflow,
  author = {{ByteDance}},
  title = {DeerFlow: An Open-Source SuperAgent Harness for Deep Research and Task Automation},
  year = {2026},
  url = {https://github.com/bytedance/deer-flow},
  note = {Accessed: 2026-03-13}
}

@misc{openclaw2026openclaw,
  author = {{OpenClaw}},
  title = {OpenClaw: Open-Source Autonomous AI Agent Framework},
  year = {2026},
  url = {https://github.com/openclaw/openclaw},
  note = {Accessed: 2026-03-13}
}

@article{liu2025deepseek,
  title={Deepseek-v3.2: Pushing the frontier of open large language models},
  author={Liu, Aixin and Mei, Aoxue and Lin, Bangcai and Xue, Bing and Wang, Bingxuan and Xu, Bingzheng and Wu, Bochao and Zhang, Bowei and Lin, Chaofan and Dong, Chen and others},
  journal={arXiv preprint arXiv:2512.02556},
  year={2025}
}

@misc{deepseekai2026deepseekv4,
      title={DeepSeek-V4: Towards Highly Efficient Million-Token Context Intelligence},
      author={DeepSeek-AI},
      year={2026},
}

@misc{cicpa_official,
  title        = {Chinese Institute of Certified Public Accountants},
  author       = {{Chinese Institute of Certified Public Accountants}},
  year         = {2026},
  howpublished = {\url{https://www.cicpa.org.cn/introcicpa/}},
  note         = {Accessed: 2026-05-15}
}

@misc{acca_official,
  title        = {Association of Chartered Certified Accountants (ACCA)},
  author       = {{Association of Chartered Certified Accountants}},
  year         = {2026},
  howpublished = {\url{https://www.accaglobal.com/gb/en.html}},
  note         = {Accessed: 2026-05-15}
}

@misc{cfa_institute,
  title        = {CFA Institute},
  author       = {{CFA Institute}},
  year         = {2026},
  howpublished = {\url{https://www.cfainstitute.org/}},
  note         = {Accessed: 2026-05-15}
}

@article{phan2025humanity,
  title={Humanity's last exam},
  author={Phan, Long and Gatti, Alice and Han, Ziwen and Li, Nathaniel and Hu, Josephina and Zhang, Hugh and Zhang, Chen Bo Calvin and Shaaban, Mohamed and Ling, John and Shi, Sean and others},
  journal={arXiv preprint arXiv:2501.14249},
  year={2025}
}

@inproceedings{mialon2023gaia,
  title={Gaia: a benchmark for general ai assistants},
  author={Mialon, Gr{\'e}goire and Fourrier, Cl{\'e}mentine and Wolf, Thomas and LeCun, Yann and Scialom, Thomas},
  booktitle={The Twelfth International Conference on Learning Representations},
  year={2023}
}

@article{Wei2025BrowseCompAS,
  title={BrowseComp: A Simple Yet Challenging Benchmark for Browsing Agents},
  author={Jason Wei and Zhiqing Sun and Spencer Papay and Scott McKinney and Jeff Han and Isa Fulford and Hyung Won Chung and Alexandre Passos and William Fedus and Amelia Glaese},
  journal={ArXiv},
  year={2025},
  volume={abs/2504.12516},
  url={https://api.semanticscholar.org/CorpusID:277857238}
}

@article{Gupta2026DeepSearchQABT,
  title={DeepSearchQA: Bridging the Comprehensiveness Gap for Deep Research Agents},
  author={Nikita Gupta and Riju Chatterjee and Lukas Haas and Connie Tao and Andrew Wang and Chang Liu and Hidekazu Oiwa and Elena Gribovskaya and Jan Ackermann and John Blitzer and Sasha Goldshtein and Dipanjan Das},
  journal={ArXiv},
  year={2026},
  volume={abs/2601.20975},
  url={https://api.semanticscholar.org/CorpusID:283897826}
}

@misc{abaskohi2026drbenchrealisticbenchmarkenterprise,
      title={DRBench: A Realistic Benchmark for Enterprise Deep Research}, 
      author={Amirhossein Abaskohi and Tianyi Chen and Miguel Muñoz-Mármol and Curtis Fox and Amrutha Varshini Ramesh and Étienne Marcotte and Xing Han Lù and Nicolas Chapados and Spandana Gella and Peter West and Giuseppe Carenini and Christopher Pal and Alexandre Drouin and Issam H. Laradji},
      year={2026},
      eprint={2510.00172},
      archivePrefix={arXiv},
      primaryClass={cs.CL},
      url={https://arxiv.org/abs/2510.00172}, 
}

@article{Du2025DeepResearchBA,
  title={DeepResearch Bench: A Comprehensive Benchmark for Deep Research Agents},
  author={Mingxuan Du and Benfeng Xu and Chiwei Zhu and Xiaorui Wang and Zhendong Mao},
  journal={ArXiv},
  year={2025},
  volume={abs/2506.11763},
  url={https://api.semanticscholar.org/CorpusID:279391682}
}

@inproceedings{Yao2025DrBA,
  title={Dr. Bench: A Multidimensional Evaluation for Deep Research Agents, from Answers to Reports},
  author={Yang Yao and Yixu Wang and Yuxuan Zhang and Yi Lu and Tianle Gu and Lingyu Li and Dingyi Zhao and Keming Wu and Haozhe Wang and Ping Nie and Yan Teng and Yingchun Wang},
  year={2025},
  url={https://api.semanticscholar.org/CorpusID:281725033}
}

@misc{wang2025liveresearchbenchlivebenchmarkusercentric,
      title={LiveResearchBench: A Live Benchmark for User-Centric Deep Research in the Wild}, 
      author={Jiayu Wang and Yifei Ming and Riya Dulepet and Qinglin Chen and Austin Xu and Zixuan Ke and Frederic Sala and Aws Albarghouthi and Caiming Xiong and Shafiq Joty},
      year={2025},
      eprint={2510.14240},
      archivePrefix={arXiv},
      primaryClass={cs.AI},
      url={https://arxiv.org/abs/2510.14240}, 
}

@misc{wan2025deepresearcharenaexamllms,
      title={DeepResearch Arena: The First Exam of LLMs' Research Abilities via Seminar-Grounded Tasks}, 
      author={Haiyuan Wan and Chen Yang and Junchi Yu and Meiqi Tu and Jiaxuan Lu and Di Yu and Jianbao Cao and Ben Gao and Jiaqing Xie and Aoran Wang and Wenlong Zhang and Philip Torr and Dongzhan Zhou},
      year={2025},
      eprint={2509.01396},
      archivePrefix={arXiv},
      primaryClass={cs.AI},
      url={https://arxiv.org/abs/2509.01396}, 
}

@article{Sharma2025ResearchRubricsAB,
  title={ResearchRubrics: A Benchmark of Prompts and Rubrics For Evaluating Deep Research Agents},
  author={Manasi Sharma and Chen Bo Calvin Zhang and Chaithanya Bandi and Clinton Wang and Ankit Aich and Huy Nghiem and Tahseen Rabbani and Ye Htet and Brian Jang and Sumana Basu and Aishwarya H. Balwani and Denis Peskoff and Marcos Ayestaran and Sean M. Hendryx and Brad Kenstler and Bing Liu},
  journal={ArXiv},
  year={2025},
  volume={abs/2511.07685},
  url={https://api.semanticscholar.org/CorpusID:282921678}
}

@inproceedings{Zhong2026DRACOAC,
  title={DRACO: a Cross-Domain Benchmark for Deep Research Accuracy, Completeness, and Objectivity},
  author={Jia Zhong and Hao Zhang and Clare Southern and Jeremy Yang and Thomas Wang and Kooseung Jung and Shu Zhang and Denis Yarats and Johnny Ho and Jerry Ma},
  year={2026},
  url={https://api.semanticscholar.org/CorpusID:285540278}
}

@misc{han2026deerbenchmarkevaluatingdeep,
      title={DEER: A Benchmark for Evaluating Deep Research Agents on Expert Report Generation}, 
      author={Janghoon Han and Heegyu Kim and Changho Lee and Dahm Lee and Min Hyung Park and Hosung Song and Stanley Jungkyu Choi and Moontae Lee and Honglak Lee},
      year={2026},
      eprint={2512.17776},
      archivePrefix={arXiv},
      primaryClass={cs.CL},
      url={https://arxiv.org/abs/2512.17776}, 
}

@misc{ye2026miroevalbenchmarkingmultimodaldeep,
      title={MiroEval: Benchmarking Multimodal Deep Research Agents in Process and Outcome}, 
      author={Fangda Ye and Yuxin Hu and Pengxiang Zhu and Yibo Li and Ziqi Jin and Yao Xiao and Yibo Wang and Lei Wang and Zhen Zhang and Lu Wang and Yue Deng and Bin Wang and Yifan Zhang and Liangcai Su and Xinyu Wang and He Zhao and Chen Wei and Qiang Ren and Bryan Hooi and An Bo and Shuicheng Yan and Lidong Bing},
      year={2026},
      eprint={2603.28407},
      archivePrefix={arXiv},
      primaryClass={cs.AI},
      url={https://arxiv.org/abs/2603.28407}, 
}

@misc{huang2026mmdeepresearchbenchbenchmarkmultimodaldeep,
      title={MMDeepResearch-Bench: A Benchmark for Multimodal Deep Research Agents}, 
      author={Peizhou Huang and Zixuan Zhong and Zhongwei Wan and Donghao Zhou and Samiul Alam and Xin Wang and Zexin Li and Zhihao Dou and Li Zhu and Jing Xiong and Chaofan Tao and Yan Xu and Dimitrios Dimitriadis and Tuo Zhang and Mi Zhang},
      year={2026},
      eprint={2601.12346},
      archivePrefix={arXiv},
      primaryClass={cs.CV},
      url={https://arxiv.org/abs/2601.12346}, 
}

@misc{zeng2026visiondeepresearchbenchmarkrethinkingvisual,
      title={Vision-DeepResearch Benchmark: Rethinking Visual and Textual Search for Multimodal Large Language Models}, 
      author={Yu Zeng and Wenxuan Huang and Zhen Fang and Shuang Chen and Yufan Shen and Yishuo Cai and Xiaoman Wang and Zhenfei Yin and Lin Chen and Zehui Chen and Shiting Huang and Yiming Zhao and Xu Tang and Yao Hu and Philip Torr and Wanli Ouyang and Shaosheng Cao},
      year={2026},
      eprint={2602.02185},
      archivePrefix={arXiv},
      primaryClass={cs.CV},
      url={https://arxiv.org/abs/2602.02185}, 
}

@article{Hu2025FinSearchCompTA,
  title={FinSearchComp: Towards a Realistic, Expert-Level Evaluation of Financial Search and Reasoning},
  author={Liang Hu and Jianpeng Jiao and Jiashuo Liu and Yanle Ren and Zhoufutu Wen and Kaiyuan Zhang and Xuanliang Zhang and Xiang Gao and Tianci He and Fei Hu and Yali Liao and Zaiyuan Wang and Chenghao Yang and Qianyu Yang and Mingren Yin and Zhiyuan Zeng and Ge Zhang and Xinyi Zhang and Xiying Zhao and Zhenwei Zhu and Hongseok Namkoong and Wenhao Huang and Yuwen Tang},
  journal={ArXiv},
  year={2025},
  volume={abs/2509.13160},
  url={https://api.semanticscholar.org/CorpusID:281325515}
}

@article{Zeng2025FinGAIAAC,
  title={FinGAIA: A Chinese Benchmark for AI Agents in Real-World Financial Domain},
  author={Lingfeng Zeng and Fangqi Lou and Zixuan Wang and Jiajie Xu and Jinyi Niu and Mengping Li and Yifan Dong and Qi Qi and Wei Zhang and Ziwei Yang and Jun Han and Ruilun Feng and Ruiqi Hu and Lejie Zhang and Zhengbo Feng and Yicheng Ren and Xin Guo and Zhaowei Liu and Dongpo Cheng and Weige Cai and Liwen Zhang},
  journal={ArXiv},
  year={2025},
  volume={abs/2507.17186},
  url={https://api.semanticscholar.org/CorpusID:280220023}
}

@article{Jin2025FinRptDE,
  title={FinRpt: Dataset, Evaluation System and LLM-based Multi-agent Framework for Equity Research Report Generation},
  author={Song Jin and Shuqi Li and Shukun Zhang and Rui Yan},
  journal={ArXiv},
  year={2025},
  volume={abs/2511.07322},
  url={https://api.semanticscholar.org/CorpusID:282911316}
}

@article{Li2026FinDeepForecastAL,
  title={FinDeepForecast: A Live Multi-Agent System for Benchmarking Deep Research Agents in Financial Forecasting},
  author={Xiangyu Li and Xuan Yao and Guohao Qi and Fengbin Zhu and Kelvin J.L. Koa and Xiang Yao Ng and Ziyang Liu and Xingyu Ni and Chang Liu and Yonghui Yang and Yang Zhang and Wenjie Wang and Fuli Feng and Chao Wang and Huanbo Luan and Xiaofen Xing and Xiangmin Xu and Tat-Seng Chua and Ke-wei Huang},
  journal={ArXiv},
  year={2026},
  volume={abs/2601.05039},
  url={https://api.semanticscholar.org/CorpusID:284544549}
}

@article{Sun2025FinResearchBenchAL,
  title={FinResearchBench: A Logic Tree based Agent-as-a-Judge Evaluation Framework for Financial Research Agents},
  author={Rui Sun and Zuo Bai and Wentao Zhang and Yuxiang Zhang and Li Zhao and Shangxue Sun and Zhengwen Qiu},
  journal={Proceedings of the 6th ACM International Conference on AI in Finance},
  year={2025},
  url={https://api.semanticscholar.org/CorpusID:280416955}
}
